\documentclass[preprintnumbers,10pt,nofootinbib]{revtex4}
\pdfoutput=1

\usepackage{amsmath,latexsym,amssymb,amsfonts}
\usepackage[pdftex]{color,graphicx}
\usepackage{bm}
\begin{document}

%\preprint{0000-00-00}

\title{\textbf{Fuzzy Euclidean wormholes in de Sitter space}}

\author{
\textsc{Pisin Chen$^{a,b,c,d}$}\footnote{{\tt pisinchen{}@{}phys.ntu.edu.tw}},
\textsc{Yao-Chieh Hu$^{a,b,c,e}$}\footnote{{\tt r04244003{}@{}ntu.edu.tw}}
and
\textsc{Dong-han Yeom$^{a}$}\footnote{{\tt innocent.yeom{}@{}gmail.com}}
}

\affiliation{
$^{a}$\small{Leung Center for Cosmology and Particle Astrophysics, National Taiwan University, Taipei 10617, Taiwan}\\
$^{b}$\small{Department of Physics, National Taiwan University, Taipei 10617, Taiwan}\\
$^{c}$\small{Graduate Institute of Astrophysics, National Taiwan University, Taipei 10617, Taiwan}\\
$^{d}$\small{Kavli Institute for Particle Astrophysics and Cosmology, SLAC National Accelerator Laboratory, Stanford University, Stanford, California 94305, USA}\\
$^{e}$\small{Department of Physics, Stockholm University, AlbaNova University Center, 106 91 Stockholm, Sweden}
}

\begin{abstract}
We investigate Euclidean wormholes in Einstein gravity with a massless scalar field in de Sitter space. Euclidean wormholes are possible due to the analytic continuation of the time as well as complexification of fields, where we need to impose the classicality after the Wick-rotation to the Lorentzian signatures. For some parameters, wormholes are preferred than Hawking-Moss instantons, and hence wormholes can be more fundamental than Hawking-Moss type instantons. Euclidean wormholes can be interpreted in three ways: (1) classical big bounce, (2) either tunneling from a small to a large universe or a creation of a collapsing and an expanding universe from nothing, and (3) either a transition from a contracting to a bouncing phase or a creation of two expanding universes from nothing. These various interpretations shed some light on challenges of singularities. In addition, these will help to understand tensions between various kinds of quantum gravity theories.
\end{abstract}

\maketitle

\newpage

\tableofcontents

\section{Introduction}

One interesting task of modern cosmology is to understand the problem of the initial singularity \cite{Hawking:1969sw,Borde:2001nh}. There are various ideas and possibilities regarding the initial singularity of our universe. It may be possible to summarize these ideas by two technical approaches. One approach is to assume that the initial singularity of our universe is describable by the wave function of the universe \cite{DeWitt:1967yk}; this approach is commonly referred to as \textit{quantum cosmology}. Traditional minisuperspace models \cite{Vilenkin:1987kf}, including the Hartle-Hawking wave function \cite{Hartle:1983ai} and loop quantum cosmology \cite{Ashtekar:2006wn}, belong to this line of philosophy. The other approach is to assume that the initial singularity of our universe is avoidable through some gravity or matter contributions. This second approach is largely motivated by quantum gravity \cite{Brandenberger:1988aj}, modified gravity \cite{Khoury:2001wf}, or modified matter field models \cite{Cai:2007qw}.

Invoking these approaches, one can imagine two possible resolutions. The first possible resolution is that the universe was \textit{never} singular and it remained approximately classical due to some higher order quantum corrections to the classical gravity. The ``big bounce'' scenario \cite{Ashtekar:2006wn,Brandenberger:1988aj,Khoury:2001wf,Cai:2007qw} can be classified as a member of this category. The second possible resolution is that the universe does reach a singularity when tracking the universe backward in time and that era maybe describable by a wave function. In this second approach, there is no definitive notion of spacetime during that era, as a consequence the universe can only be described probabilistically \cite{DeWitt:1967yk,Vilenkin:1987kf,Hartle:1983ai}.

It is interesting to note that traditionally the Euclidean path integral \cite{Gibbons:1994cg} and the Hartle-Hawking wave function approach \cite{Hartle:1983ai} prefer the latter resolution, while loop quantum gravity \cite{Ashtekar:2006wn} prefers the big bounce scenario\footnote{There would be a way to assign a probability for each history in loop quantum cosmology, but at once an observer is in a certain history, one may not need to care about the other histories. This would be the reason why lots of authors do not care about the probability of a bouncing universe.}. This means that following the Euclidean path integral approach, one deals with the superposition of various histories \cite{Hartle:2015bna,Hwang:2012mf}, whereas following the loop quantum cosmology philosophy, one would only deal with a unique history for both classical and quantum regimes \cite{Ashtekar:2015dja}. However, if two approaches are on the same line of the quantum gravity, then there should be the ground of common understanding on the resolution of the singularity.

In order to find a clue, in this paper, we start from the Hartle-Hawking wave function with a scalar field in de Sitter space \cite{Hartle:2008ng}. Traditionally, in the Euclidean quantum cosmology, people investigate compact instantons, the so-called Hawking-Moss type instantons \cite{Hawking:1981fz}. In this paper, we generalize these instantons to \textit{non-compact instantons}, where the complexification of fields is naturally allowed. This kind of non-compact instantons can be named as Euclidean wormholes \cite{Hawking:1988ae}. In addition, due to the nature of the complexified fields, it is fair to refer to these as \textit{fuzzy} Euclidean wormholes (similar efforts are in \cite{Twamley:1992hu}).

After classifying various properties of fuzzy Euclidean wormholes, we will show that these Euclidean wormholes can be understood in \textit{both} ways: either as a classical or quantum bouncing process (compatible with loop quantum cosmology or big bounce models) or a creation of two universes from nothing (compatible with minisuperspace quantum cosmology approaches) \cite{Robles-Perez:2013kva}. We note that these interpretations are achieved by slightly extending our complexified instantons from compact to non-compact ones. Our new interpretation can help to shed light on the Euclidean path integral approach as well as various big bounce models including loop quantum cosmology.

In SEC.~\ref{sec:euc}, we discuss Euclidean wormholes in the analytic level. In SEC.~\ref{sec:euccomp}, we discuss a numerical investigation of complexified wormholes and their classicalization. In SEC.~\ref{sec:prob}, we discuss probabilities of various shapes of wormholes. In SEC.~\ref{sec:phys}, we summarize physical issues relating Euclidean wormholes. Finally, in SEC.~\ref{sec:con}, we summarize our results and outline possible future issues.

\section{\label{sec:euc}Euclidean wormholes}

\subsection{The Hartle-Hawking wave function}

The Hartle-Hawking wave function \cite{Hartle:1983ai}, or the ground state wave function of the universe is described by
\begin{eqnarray}
\Psi[h_{\mu\nu}, \chi] = \int_{\partial g = h, \partial \phi = \chi} \mathcal{D}g\mathcal{D}\phi \;\; e^{-S_{\mathrm{E}}[g,\phi]},
\end{eqnarray}
where $h_{\mu\nu}$ and $\chi$ are the boundary values of the Euclidean metric $g_{\mu\nu}$ and the matter field $\phi$. This will be approximated by steepest-descents, or equivalently, the Euclidean on-shell histories\footnote{Recently, there is a discussion that there would be a better approximation method of the Lorentzian path integral \cite{Feldbrugge:2017kzv}. Applications for the Lorentzian path integral by using the Picard-Lefschetz theory (rather than the instanton method) would be an interesting topic and we leave this for future investigations.}. In general, due to the Wick-rotations, we require that all functions should be complex-valued. These complex-valued instantons are called by \textit{fuzzy instantons} \cite{Hartle:2008ng}.

However, not all the fuzzy instantons are relevant for the creation of universes. After the long Lorentzian time, the manifold should be smoothly connected to the observer, where the observer is assumed to be classical. If we approximately write the wave function (using the steepest-descent approximation) as
\begin{equation}
\Psi[q_{I}] \simeq A[q_{I}] e^{i S[q_{I}]},
\end{equation}
where $q_{I}$ are canonical variables with $I=1,2,3, ...$, then the \textit{classicality} means that
\begin{equation} \label{eqn:classicality}
\left|\nabla_I A\left[q_{I}\right]\right|\ll \left|\nabla_I S\left[q_{I}\right]\right|,
\end{equation}
for all $I$. Then this history satisfies the semi-classical Hamilton-Jacobi equation. In order to check the classicality formally, we need to compare with instantons by varying the boundary values. For practical purposes, the intuitive meaning of the classicality is that when we solve on-shell Euclidean equations, although we introduce complex-valued functions, such \textit{complex-valued functions should approach to real valued functions} on the boundary. %after the Wick-rotation and after a sufficient Lorentzian time.

In this paper, we consider the following action
\begin{eqnarray}
S = \int \sqrt{-g}dx^{4} \left[ \frac{R}{16\pi} - \frac{1}{2} \left(\nabla \phi\right)^{2} - V(\phi) \right].
\end{eqnarray}
In addition, we give the Euclidean minisuperspace metric as
\begin{eqnarray}
ds^{2}_{\mathrm{E}} = d\tau^{2} + a^{2}(\tau) d\Omega_{3}^{2}.
\end{eqnarray}
Then the equations of motion that we have to satisfy are as follows:
\begin{eqnarray}
\dot{a}^{2} - 1 - \frac{8\pi a^{2}}{3} \left( \frac{\dot{\phi}^{2}}{2} - V \right) &=& 0,\\
\ddot{\phi} + 3 \frac{\dot{a}}{a} \dot{\phi} - V' &=& 0, \\
\frac{\ddot{a}}{a} + \frac{8\pi}{3}\left( \dot{\phi}^{2} + V \right) &=& 0.
\end{eqnarray}

\subsection{Approximate analytic solution in Euclidean section}

In order to demonstrate a solution, we first construct an approximate Euclidean wormhole solution that can be calculated by an analytic way.

If the potential is flat, $V(\phi) = V_{0}$, then
\begin{eqnarray}
\frac{\ddot{\phi}}{\dot{\phi}} = - 3 \frac{\dot{a}}{a},
\end{eqnarray}
and hence
\begin{eqnarray}
\dot{\phi} = \mathcal{A} a^{-3}
\end{eqnarray}
with a constant $\mathcal{A}$. The equation for $a$ becomes
\begin{eqnarray}
\dot{a}^{2} + V_{\mathrm{eff}}(a) &=& 0,\\
V_{\mathrm{eff}}(a) &=& - 1 - \frac{8\pi}{3} \left( \frac{\mathcal{A}^{2}}{2a^{4}} - V_{0} a^{2} \right).
\end{eqnarray}
Hence, $V_{\mathrm{eff}}(a) < 0$ is the physically allowed region.

For our own interests, we set $\mathcal{A} = i \mathcal{B}$ with a real value $\mathcal{B}$. Then the effective potential becomes
\begin{eqnarray}\label{eq:approx}
V_{\mathrm{eff}}(a) = - 1 + \left( \frac{a_{0}^{4}}{a^{4}} + \epsilon \frac{a^{2}}{\ell^{2}} \right),
\end{eqnarray}
where $a_{0}=(4\pi \mathcal{B}^{2}/3)^{1/4}$, $\ell = (3/8\pi \left|V_{0}\right|)^{1/2}$, and $\epsilon = \mathrm{sign} V_{0}$. Here, $\mathcal{B}$ should not depend on $a$.

If $\epsilon > 0$, then there can be two solutions for $V(a)=0$. If $\epsilon \leq 0$, there is one solution for $V(a)=0$. If $a_{0} \ll \ell$, then one solution is $a_{\mathrm{min}} \simeq a_{0}$ and the other solution is $a_{\mathrm{max}} \simeq \ell$ for $\epsilon = 1$ case; otherwise, we regard that $a_{\mathrm{max}} = \infty$ for $\epsilon \leq 0$. The physical solutions are allowed between $a_{\mathrm{min}} \leq a \leq a_{\mathrm{max}}$. In this paper, we mainly focus on the de Sitter case: $\epsilon = 1$.

One problem of this simple solution is that, after the Wick-rotation, there still remains the imaginary part, especially in the metric part. Therefore, we need more numerical investigations to check whether such a classicalization is indeed possible or not.

\subsection{Comparison to other models}

There can be various models that realize Euclidean wormholes. In this section, we illustrate three interesting examples from quantum gravity, modified matter, and modified gravity models.

\paragraph{Quantum gravity: excitations} The Wheeler-DeWitt equation of Einstein gravity with a conformally coupled scalar field $\phi$ can be presented as follows \cite{Hartle:1983ai}:
\begin{eqnarray}
\frac{1}{2} \left[ - \frac{1}{a^{p}} \frac{d}{da} \left( a^{p} \frac{d}{da} \right)  + \left(a^{2} - \frac{\Lambda}{3}a^{4} \right) \right] \psi(a) &=& E \psi(a),\\
\frac{1}{2} \left(- \frac{d^{2}}{d\chi^{2}} + \chi^{2} \right) &=& E \zeta(\chi),
\end{eqnarray}
where $\Psi(a,\chi) = \psi(a) \zeta(\chi)$, $\chi = (4\pi/3)^{1/2}a\phi$, and $p$ is a constant for denoting a different factor ordering. Here, the separation variable $E$ should be quantized: $E_{n} = n+1/2$. Therefore, in fact, the ground state is $E_{0} = 1/2$ (while in the original paper of Hartle and Hawking, they inserted a constant $\epsilon = -1/2$ in order to make $E_{0} = 1/2 + \epsilon = 0$). In this case, the steepest-descent for $\psi(a)$ should satisfy \cite{Robles-Perez:2013kva}
\begin{eqnarray}\label{eq:grndst}
\dot{a}^{2} = 1 - \frac{\Lambda}{3}a^{2} - \frac{2E_{n}}{a^{2}}.
\end{eqnarray}
This equation allows Euclidean wormholes, where there are two turning points $a_{\mathrm{min}}^{2} \simeq 2E_{n}$ and $a_{\mathrm{max}}^{2} \simeq 3/\Lambda$.

\paragraph{Modified matter: string theory} Euclidean wormholes are possible with an axion-induced model \cite{Giddings:1987cg}, where this can be embedded by string theory \cite{ArkaniHamed:2007js}. The axion-induced Euclidean action becomes
\begin{eqnarray}
S = \int \sqrt{+g}d^{4}x \left[ -  \frac{R}{16\pi} + \frac{1}{2} G_{IJ}(\phi) \nabla^{\mu} \phi^{I} \nabla_{\mu} \phi^{J} \right],
\end{eqnarray}
where $I, J$ label different species of scalar fields and $G_{IJ}$ can have negative signs for axionic scalars. This looks like an apparent ghost field in Euclidean signatures, though it has the correct sign for Lorentzian signatures. We can present the dynamics of the metric by (for asymptotically flat cases)
\begin{eqnarray}
\dot{a}^{2} = 1 + \frac{C}{a^{4}},
\end{eqnarray}
where $C$ can be chosen negative due to the wrong sign of the kinetic terms of the axion-induced scalars. In this case, there appears an asymptotically flat Euclidean wormhole where the minimum value is at $a_{\mathrm{min}}^{2} = \sqrt{|C|}$.

\paragraph{Modified gravity: massive gravity} One another good motivation of Euclidean wormholes comes from massive gravity \cite{deRham:2010kj}. There are interesting examples of instantons motivated by massive gravity or bigravity models \cite{Zhang:2012ap}. For example, in the following form of the massive gravity inspired action \cite{Lin:2013aha}, we can obtain Euclidean wormholes:
\begin{eqnarray}
S_{\mathrm{E}} = - \int d^{4}x \sqrt{+g} \left(\frac{R}{2} - \Lambda + m_{1}^{2} G^{\mu\nu} f_{\mu\nu} - m_{2}^{2} \left( c_{0} + c_{1} f + c_{2} f^{2} + d_{2} f^{\mu}_{\nu} f^{\nu}_{\mu} + ... \right) \right),
\end{eqnarray}
where we give the ansatz:
\begin{eqnarray}
f_{\mu\nu} = \mathrm{diag} \left[ 0, 1, \sin^{2} \psi, \sin^{2}\psi \sin^{2}\theta \right].
\end{eqnarray}
One can rewrite the equation of motion as (see Appendix)
\begin{eqnarray}
\dot{a}^{2} = \left(1-\frac{\Lambda_{\mathrm{eff}}}{3} a^{2} - \frac{\alpha}{a^{2}}\right)  \left( 1+\frac{m_{1}^{2}}{a^{2}} \right)^{-1}.
\end{eqnarray}
If $\alpha > 0$, then $a=0$ is not allowed and hence there exists an Euclidean wormhole solution. If $\Lambda_{\mathrm{eff}} < 0$, then there is one zero
\begin{eqnarray}
a_{\mathrm{min}}^{2} = \frac{3}{\left|\Lambda_{\mathrm{eff}}\right|} \left( \frac{ - 1 + \sqrt{1 + 4 \alpha \left|\Lambda_{\mathrm{eff}}\right|/3}}{2} \right)
\end{eqnarray}
and $a_{\mathrm{min}} < a$ is the allowed region. On the other hand, if $\Lambda_{\mathrm{eff}} > 0$, then there are two zeros
\begin{eqnarray}
a_{\mathrm{min,\;max}}^{2} = \frac{3}{\Lambda_{\mathrm{eff}}} \left( \frac{1 \pm \sqrt{1 - 4 \alpha \Lambda_{\mathrm{eff}}/3}}{2} \right),
\end{eqnarray}
where we need to assume $\alpha \Lambda_{\mathrm{eff}} < 3/4$; then, $a_{\mathrm{min}} < a < a_{\mathrm{max}}$ is the allowed region, where $a_{\mathrm{min}}^{2} \simeq \alpha$ and $a_{\mathrm{max}}^{2} \simeq 1/H_{\mathrm{eff}}^{2}$.

\paragraph{Summary} To summarize, there can be various motivations of Euclidean wormholes. Especially, if there is an order $a^{-n}$ correction to the Friedman equation for the small $a$ limit, then we may obtain Euclidean wormholes (though the sign of the term is also important). From quantum gravitational excitations, there is a correction of the order of $\sim n a^{-2}$ (where $n$ is a number to characterize excited states). From string-inspired corrections, \textit{due to the wrong sign of the kinetic terms}, there is a correction on the order of $\sim a^{-4}$. From massive gravity inspired models, there is a correction of the order of $\sim m_{1}^{2}m_{2}^{2} a^{-4}$ or $\sim m_{2}^{2} a^{-2}$ (depending on the choice of model parameters $m_{1}$ and $m_{2}$). In this paper, we also obtain the $\sim a^{-4}$ term due to the wrong sign of the kinetic term motivated by the analytic continuation of the time and the complexification of fields.

\begin{figure}
\begin{center}
\includegraphics[scale=0.75]{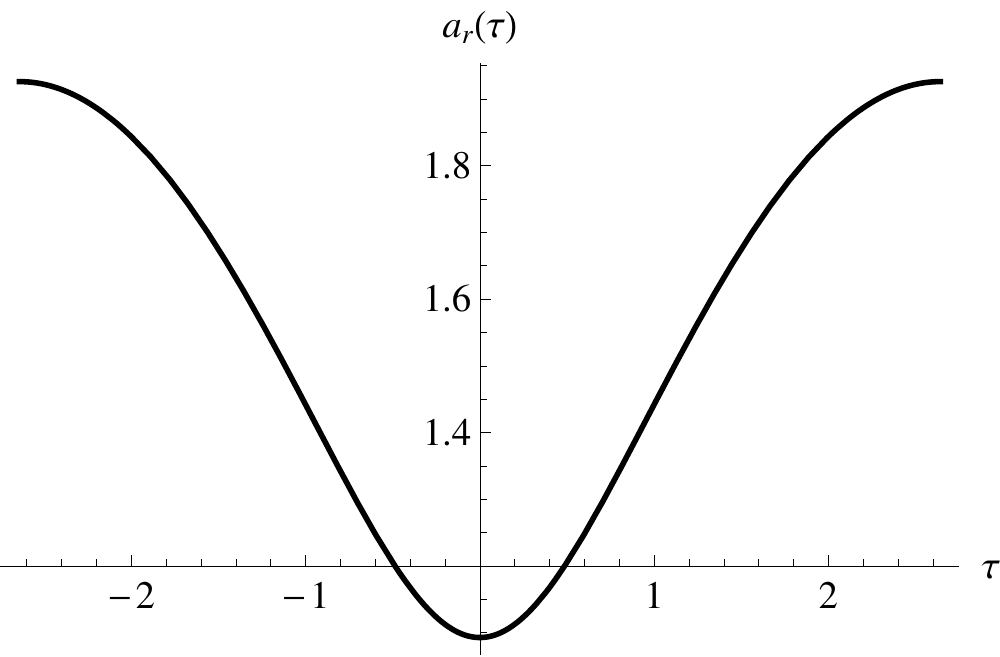}
\includegraphics[scale=0.75]{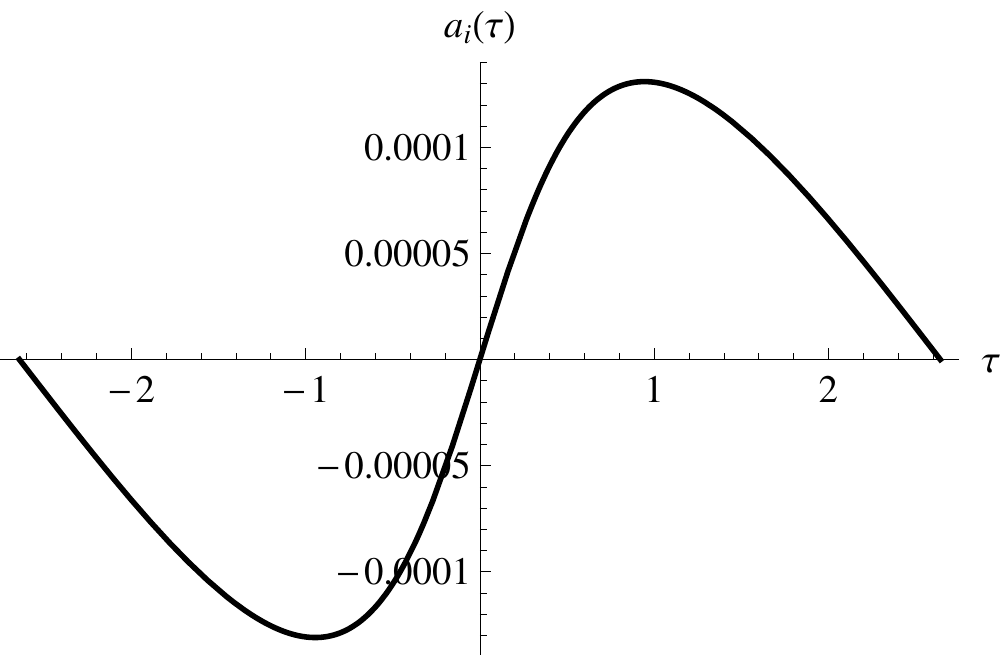}
\includegraphics[scale=0.75]{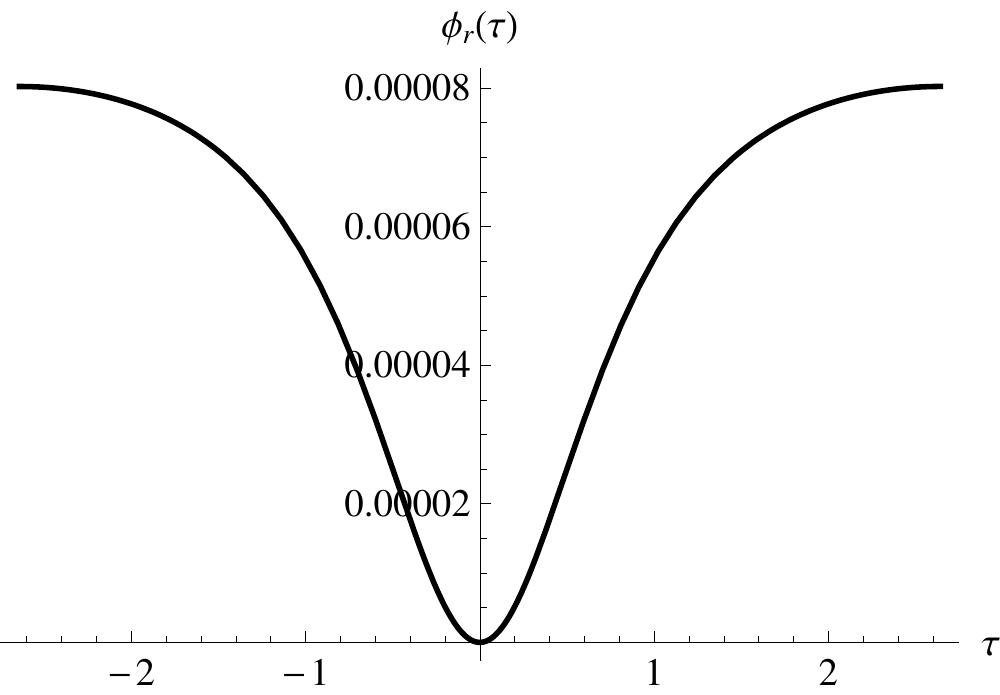}
\includegraphics[scale=0.75]{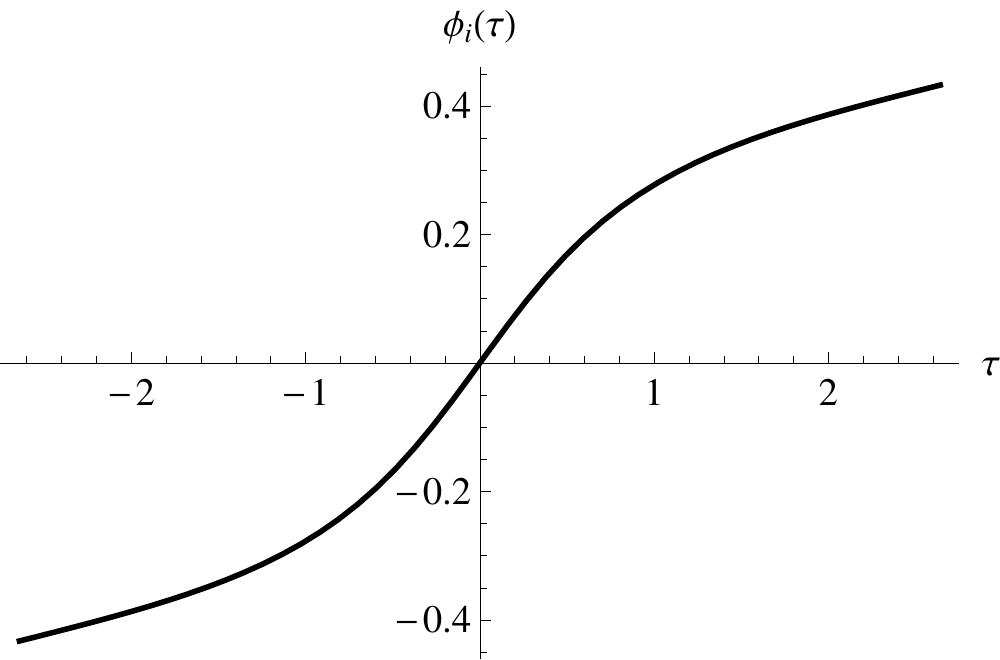}
\caption{\label{fig:exampleE}Example of a classicalized wormhole solution in Euclidean signatures ($a_{0} = 1$, $\ell = 2$, $\zeta=0.0000001$).}
%\end{center}
%\end{figure}
%\begin{figure}
%\begin{center}
\includegraphics[scale=0.75]{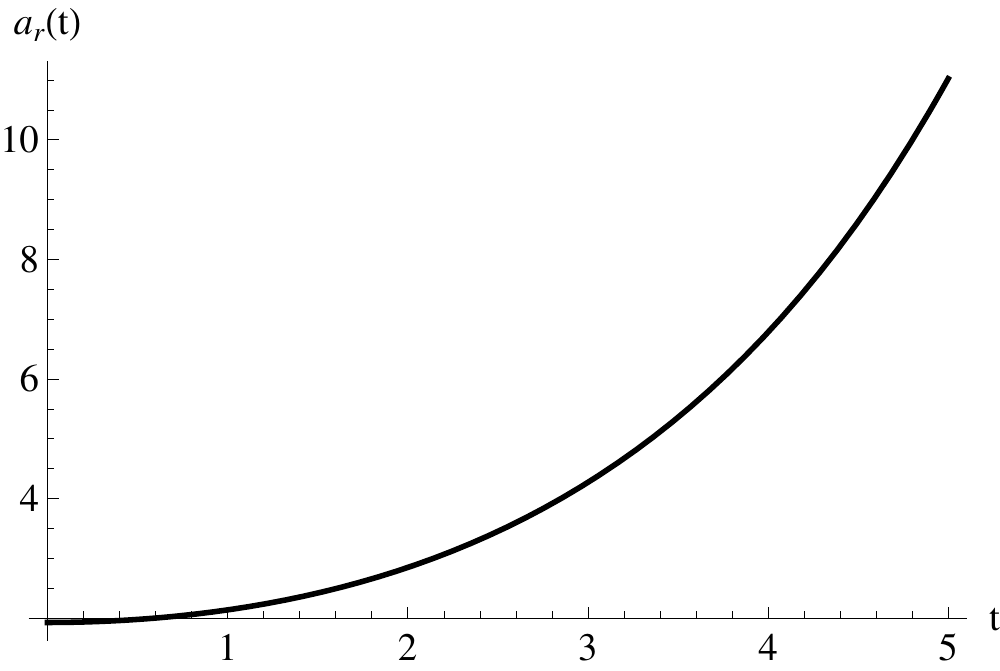}
\includegraphics[scale=0.75]{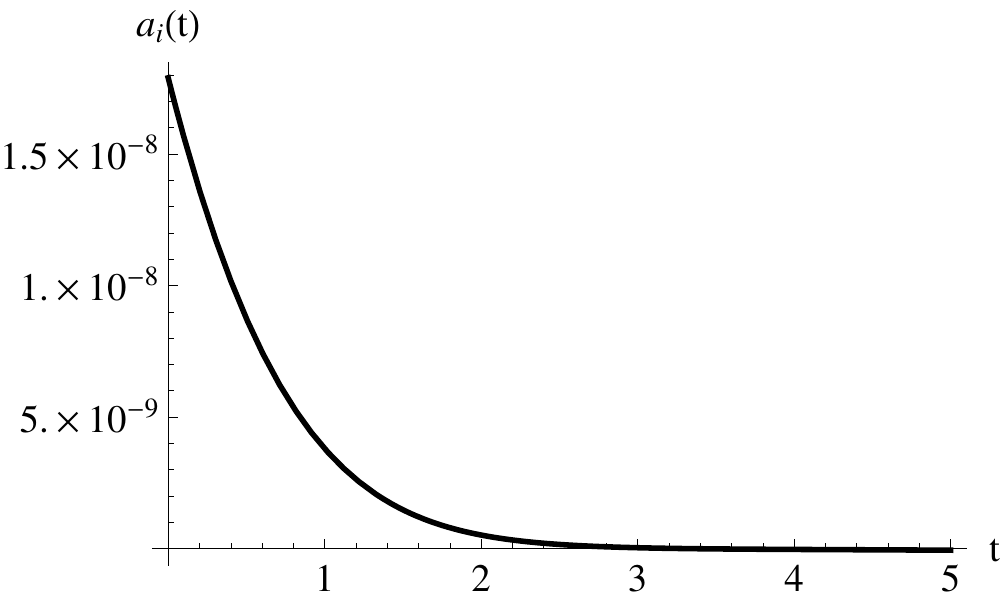}
\includegraphics[scale=0.75]{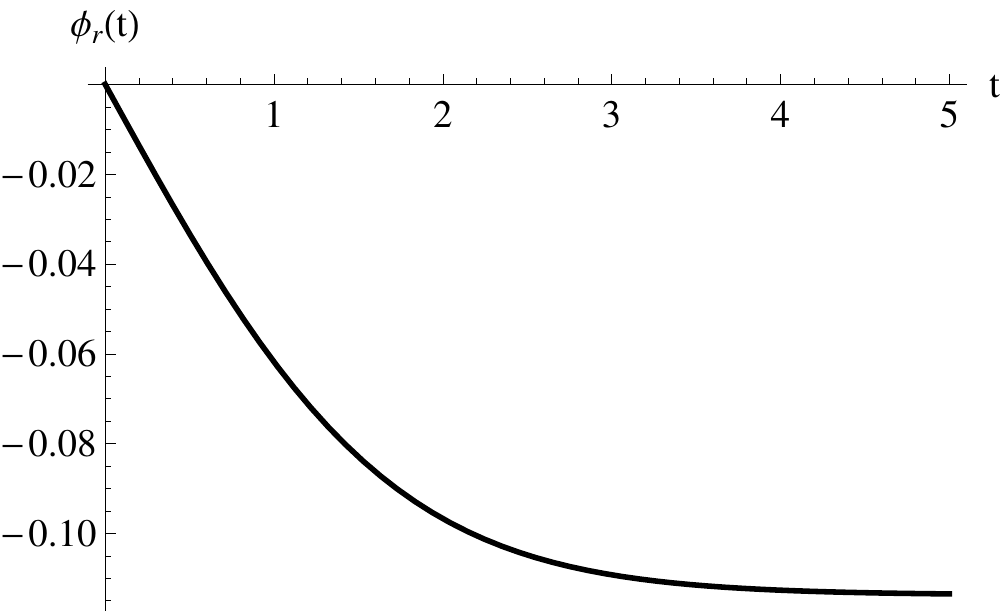}
\includegraphics[scale=0.75]{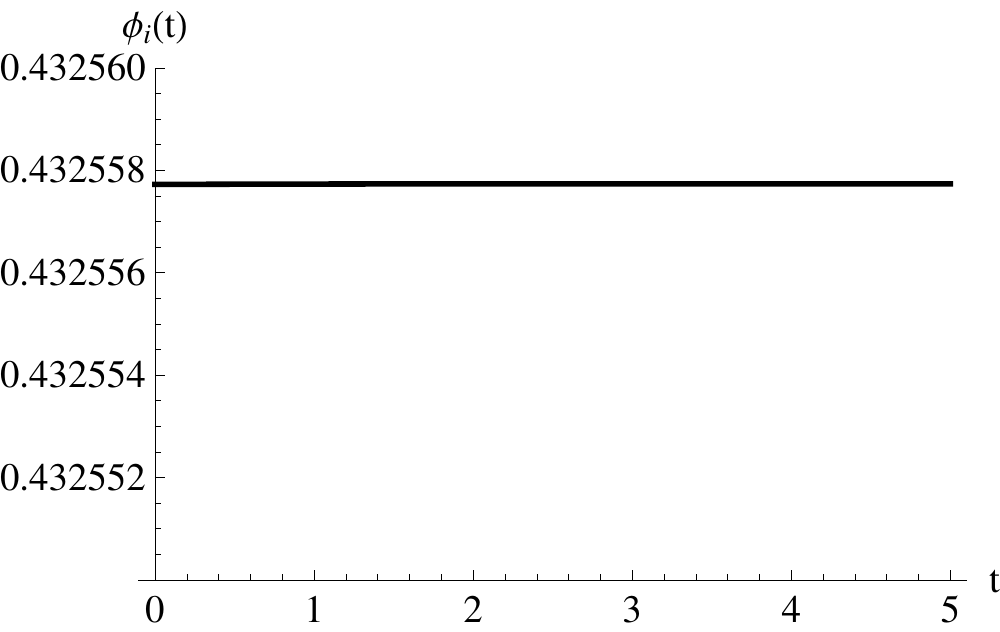}
\caption{\label{fig:exampleL}Example of a classicalized wormhole solution in Lorentzian signatures ($a_{0} = 1$, $\ell = 2$, $\zeta=0.0000001$). As $t$ increases, the scalar field approaches a constant and hence the scalar field is classicalized; at the same time, $a_{i}$ approaches to zero and hence the metric is also classicalized.}
\end{center}
\end{figure}

\begin{figure}
\begin{center}
\includegraphics[scale=0.75]{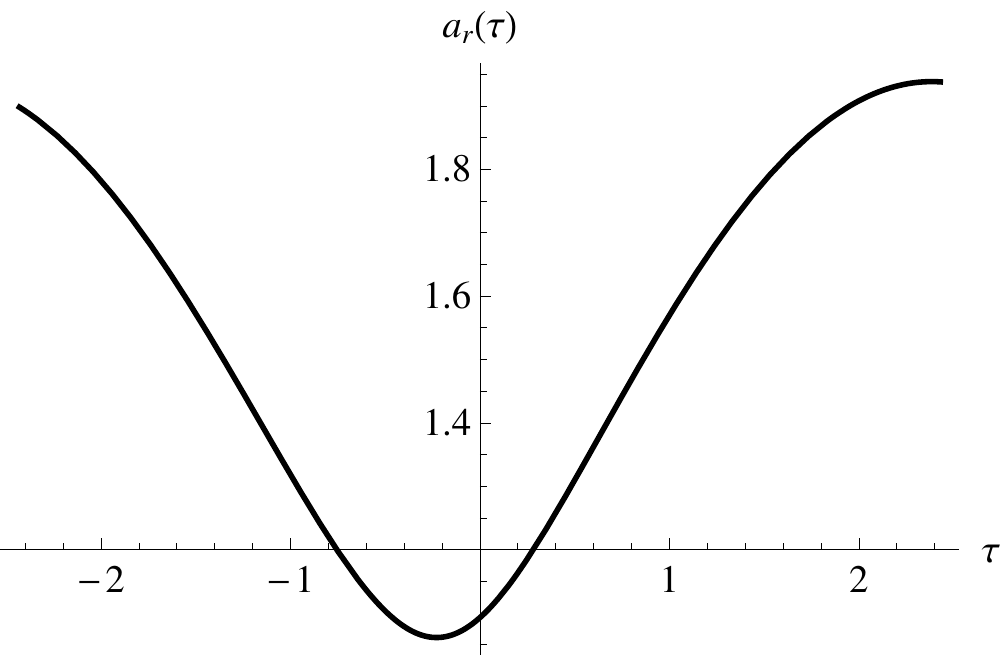}
\includegraphics[scale=0.75]{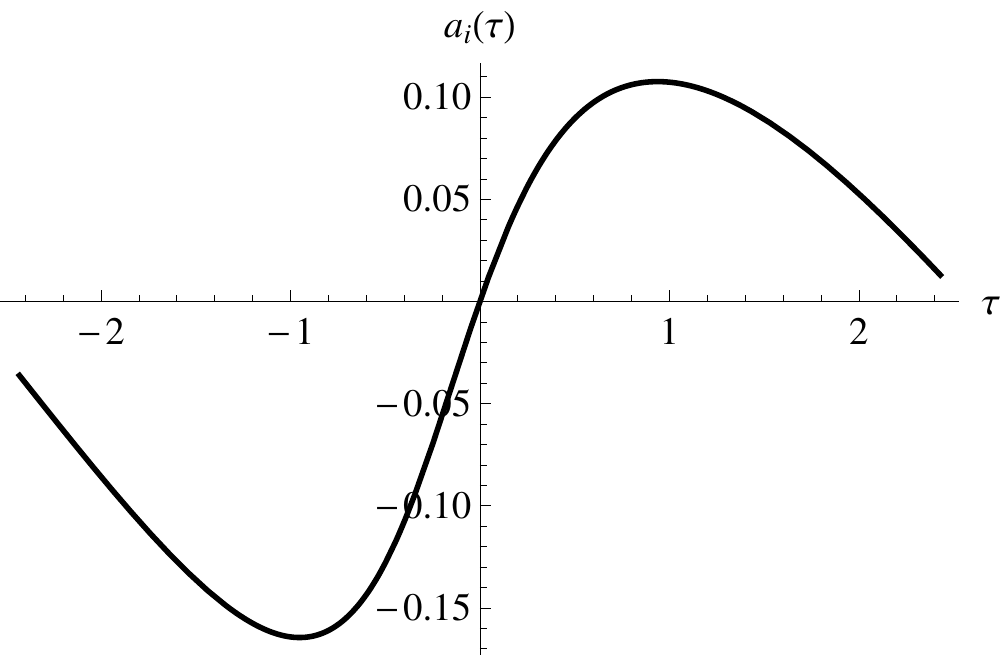}
\includegraphics[scale=0.75]{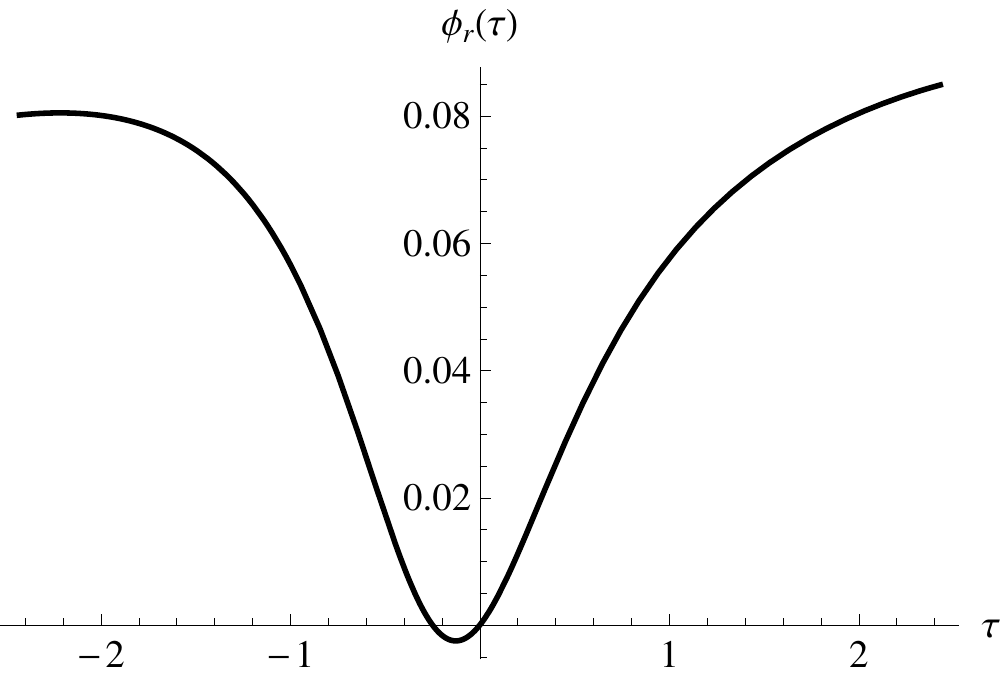}
\includegraphics[scale=0.75]{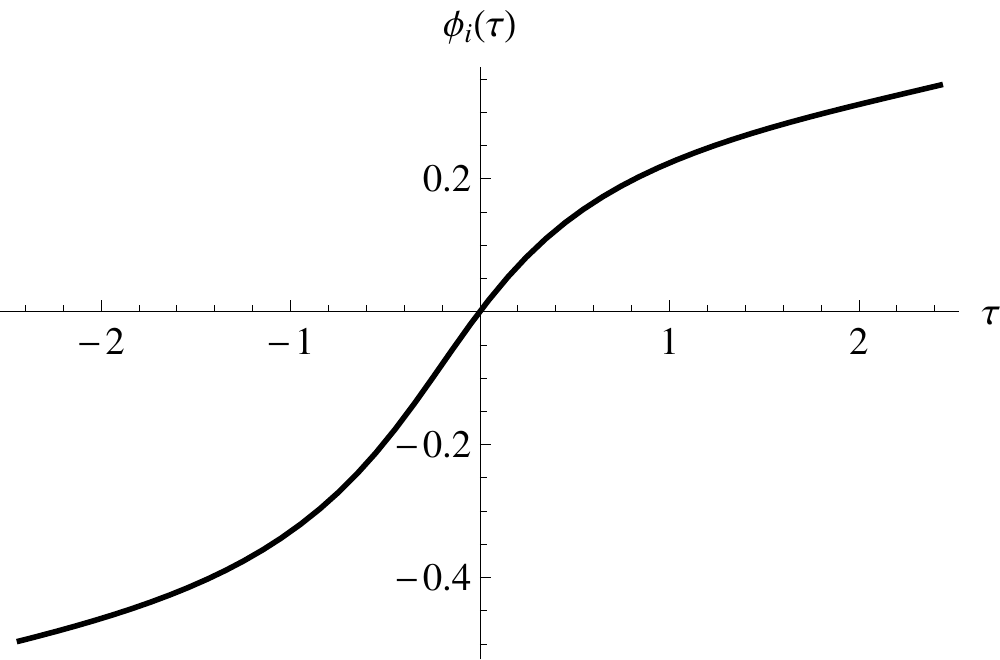}
\caption{\label{fig:exampleE2}Example of a classicalized wormhole solution in Euclidean signatures ($a_{0} = 1$, $\ell = 2$, $\zeta=0.1$).}
%\end{center}
%\end{figure}
%\begin{figure}
%\begin{center}
\includegraphics[scale=0.75]{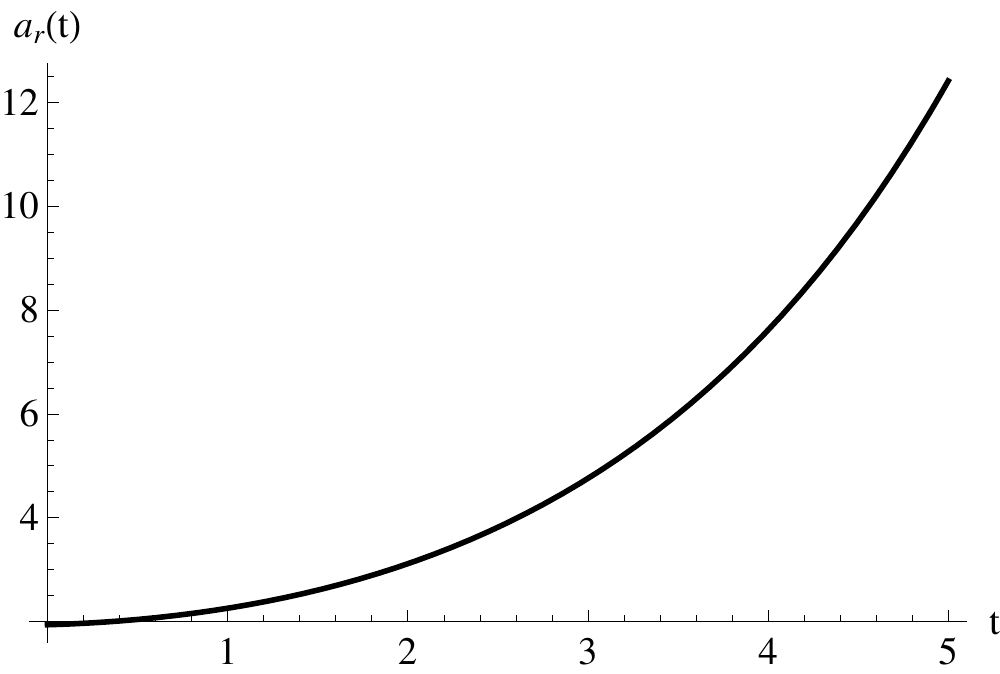}
\includegraphics[scale=0.75]{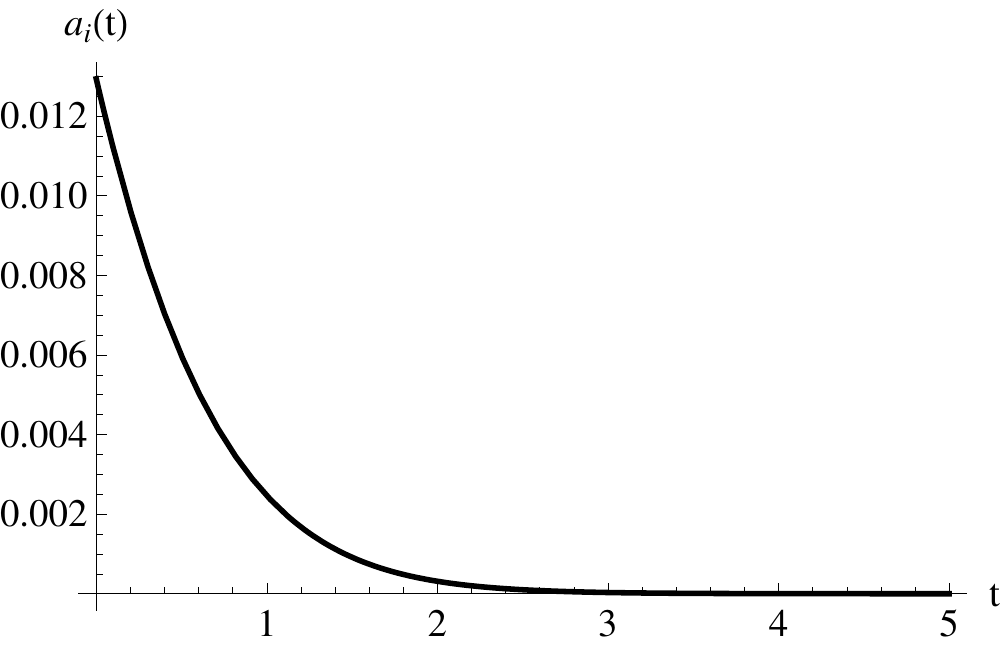}
\includegraphics[scale=0.75]{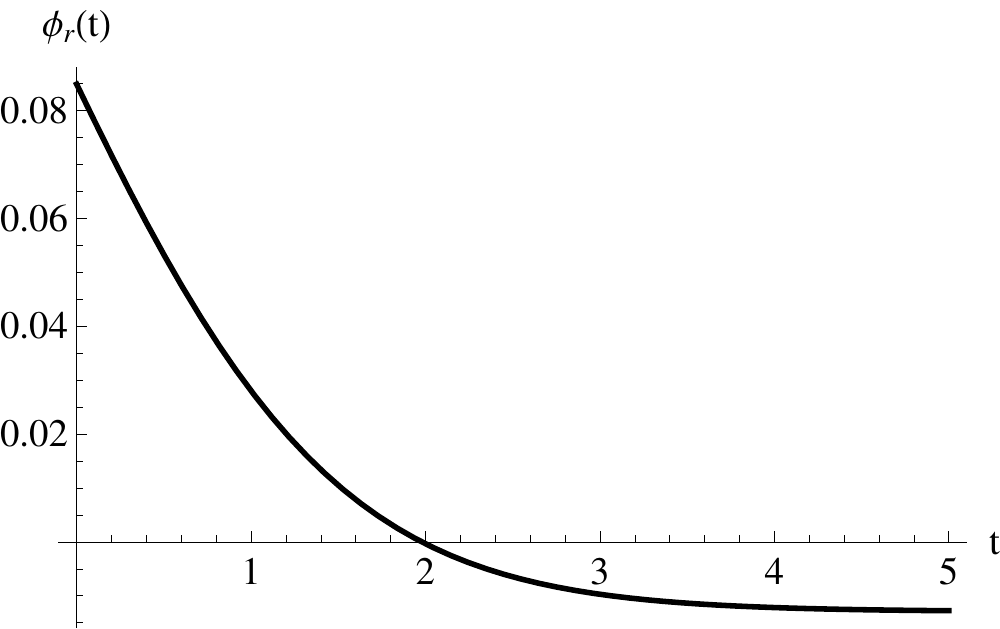}
\includegraphics[scale=0.75]{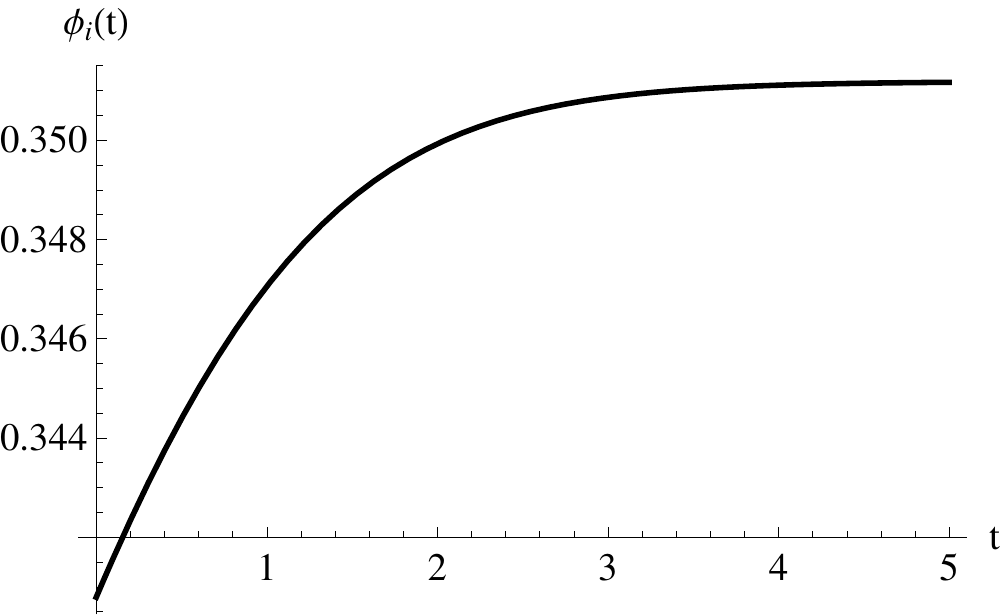}
\caption{\label{fig:exampleL2}Example of a classicalized wormhole solution in Lorentzian signatures ($a_{0} = 1$, $\ell = 2$, $\zeta=0.1$).}
\end{center}
\end{figure}

\section{\label{sec:euccomp}Euclidean complexified wormholes}

\subsection{Analytic continuation and classicalization}

The solution of the previous section is exact, but we did not apply analytic continuations to Lorentzian signatures yet. The problem is that after the Wick-rotation, the scalar field should have both of real and imaginary parts. Hence, the metric should be also complexified. In order to control the imaginary part of the metric, we choose the following initial conditions.
\begin{eqnarray}
&&a_{r}(0) = a_{\mathrm{min}},\\
&&a_{i}(0) = 0,\\
&&\dot{a}_{r}(0) = \sqrt{\frac{4\pi}{3}} \frac{\mathcal{B}}{a_{\mathrm{min}}^{2}} \sqrt{\sinh{\zeta}\cosh{\zeta}},\\
&&\dot{a}_{i}(0) = \sqrt{\frac{4\pi}{3}} \frac{\mathcal{B}}{a_{\mathrm{min}}^{2}} \sqrt{\sinh{\zeta}\cosh{\zeta}},\\
&&\phi_{r}(0) = 0,\\
&&\phi_{i}(0) = 0,\\
&&\dot{\phi}_{r}(0) = \frac{\mathcal{B}}{a_{\mathrm{min}}^{3}} \sinh{\zeta},\\
&&\dot{\phi}_{i}(0) = \frac{\mathcal{B}}{a_{\mathrm{min}}^{3}} \cosh{\zeta},
\end{eqnarray}
where $\zeta$ is a free parameter. The turning time is $X = \delta \times \tau_{\mathrm{max}}$, where $\dot{a}(\tau_{\mathrm{max}}) = 0$. Since there is no potential, the field value of $\phi$ can be chosen arbitrarily. Therefore, the only meaningful variable is $\mathcal{B}$, $\zeta$, and $\delta$. For a given $\mathcal{B}$ and $\zeta$, by tuning a free parameter $\delta$, we can find a classicalized solution \cite{Hwang:2011mp}.

FIGs.~\ref{fig:exampleE} and \ref{fig:exampleL} are the case when $a_{0} = 1$, $\ell = 2$, $\zeta = 0.0000001$, and $\delta \simeq 1.00000002199$. FIGs.~\ref{fig:exampleE2} and \ref{fig:exampleL2} are the case when $a_{0} = 1$, $\ell = 2$, $\zeta = 0.1$, and $\delta \simeq 1.01879$. Hence, as $\zeta$ increases from zero, the wormhole solution becomes asymmetric. After the Wick-rotation, each parts of the scalar field approaches a constant (due to the friction of the expanding universe). Since there is no potential, if the scalar field stops to move, then it is already classicalized. At the same time, the imaginary part of the metric approaches to zero. This completes the classicalization of the solution.

Here, we comment on the issue about whether our classicality condition is sufficient, even if the scalar field approaches a constant imaginary value rather than zero. Since there is no potential for the scalar field, there exists a shift symmetry under the transformation of the scalar field, $\phi \rightarrow \phi + \phi_{0}$, where $\phi_{0}$ is an arbitrary constant (this can be a pure imaginary number). Hence, for each (left or right) side of the wormhole, one can choose a suitable $\phi_{0}$ so that the field value asymptotically approaches zero. One possible worry is that if we choose a specific $\phi_{0}$ on one side (and make the field approach zero asymptotically), then in terms of the other side, the field value becomes a pure imaginary number. Is it still consistent to impose the classicality on the other side? Regarding this, there are two explanations. First, in terms of a one-side observer, he/she will not care about the classicality of the other side but will only care about the \textit{finiteness} of the action integral. The condition for the finiteness of the action integral corresponds to imposing the conditions $a_{i} \rightarrow 0$ and $\phi \rightarrow \mathrm{constant}$ for both sides of the wormhole. Therefore, at once the scalar field approaches a constant from each side, we can do a consistent theory at least for one side since the probability is well defined. Second, observers on one side cannot compare notes with observables on the other side. Therefore, a left-side observer can locally redefine the scalar field by choosing a constant, while a right-side observer can choose his/her own constant independently; nobody can compare their different choices unless there is an interaction term. Hence, local observers on each side can construct a suitable classical history independently. Of course, if the scalar field has interaction terms, this argument does not hold. In that case, we need more careful treatment, where we leave this for a future work\footnote{Regarding this issue, our solution satisfies classical equations of motion for both ends of the wormhole, because there is no potential term of the scalar field and there is no contribution from the kinetic term as long as the scalar field is a constant. The existence of (physically undetectable) constant imaginary scalar field value would be uncomfortable, but such a counter-intuitive but mathematically consistent effect can be appeared in the Euclidean path integral formalism. One illustrative example is a paper by Hartle, Hawking and Hertog \cite{Hartle:2012qb}. In this work, they reported that there exists a contour following a \textit{Euclidean} time direction which demonstrates a \textit{Lorentzian} de Sitter space, even with a negative cosmological constant. This is possible since the metric is \textit{pure imaginary} through the contour. Of course, by extending the model to more complicated interactions, such a fancy behavior would cause unphysical problems; perhaps, this would be the origin of the uncomfortable feeling. However, at least, in this simple model, such a pure imaginary metric satisfies the classicality condition, and hence it should be regarded as a classical history. We also argue that the same thing happens in our wormhole solutions, and hence they should be accepted as classical universes at least there is no potential term.}.

\subsection{Existence of classicalized turning time and geometrical interpretations}

\begin{figure}
\begin{center}
\includegraphics[scale=0.3]{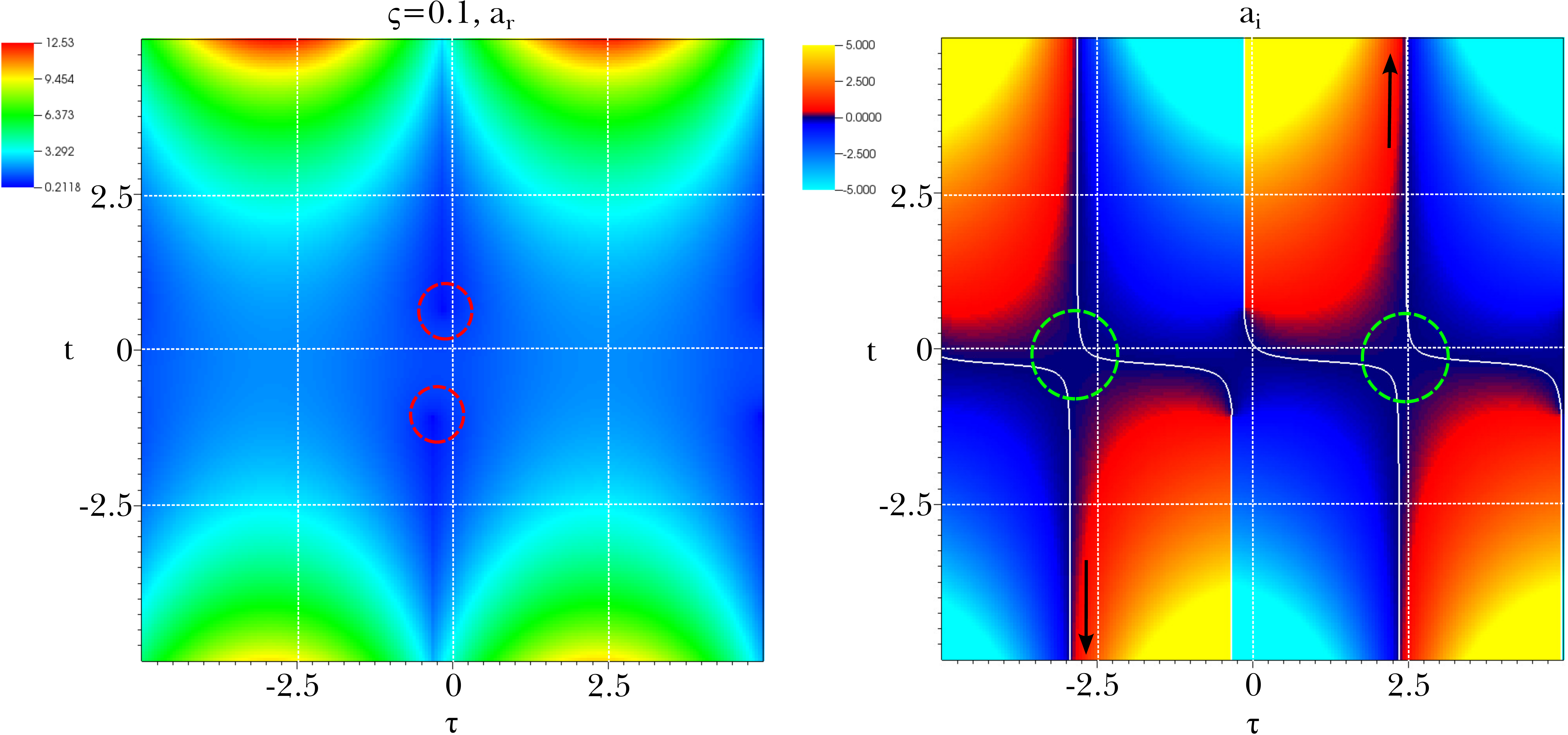}
\caption{\label{fig:zeta=01}Instanton solution as a function of the complex time plane for $a_{0} = 1$, $\ell = 2$, $\zeta=0.1$. Left is $a_{r}$ and right is $a_{i}$, where white curves of the right figure denote $a_{i}=0$. In the left figure, red circles denote poles (singularities). In the right figure, left and right circles are both of turning points of wormholes.}
\includegraphics[scale=0.3]{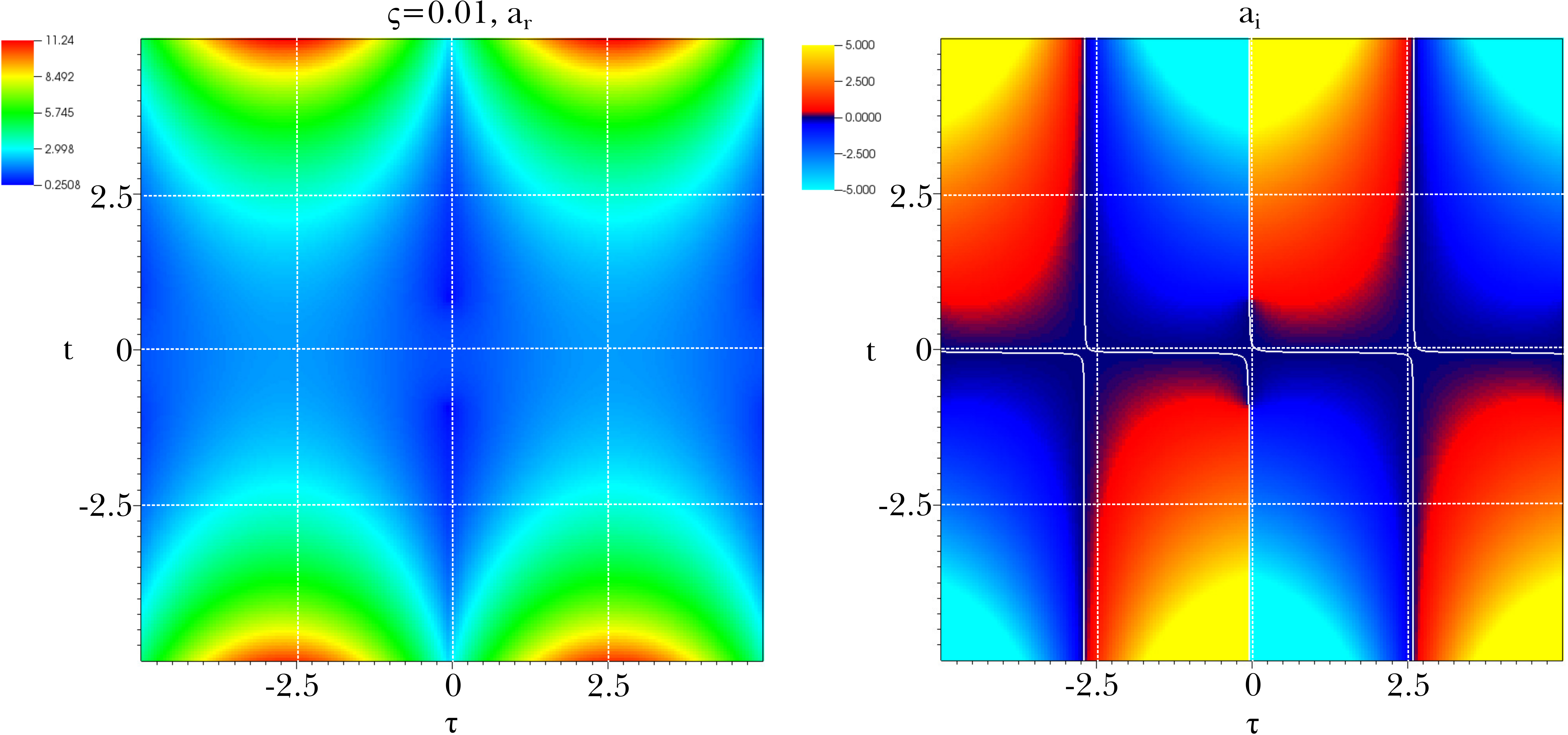}
\caption{\label{fig:zeta=001}Instanton solution as a function of the complex time plane for $a_{0} = 1$, $\ell = 2$, $\zeta=0.01$. Left is $a_{r}$ and right is $a_{i}$, where white curves of the right figure denote $a_{i}=0$.}
\end{center}
\end{figure}

In fact, we need to check the classicality for both directions of wormholes (since a wormhole has two branches). By varying initial conditions, can we sure whether there exists a classicalized time direction? This can be checked by solving the solution on the complex time plane \cite{Chen:2015ria,Battarra:2014xoa}. FIGs.~\ref{fig:zeta=01} and \ref{fig:zeta=001} are examples of $a_{0}=1$, $\ell=2$, and $\zeta = 0.1$ or $0.01$. These show typical behaviors of solutions on the complex time plane. Regarding the classicalization, we do not need to worry about scalar fields, since these will stop due to the expansion of $a_{r}$. The only function that we need to concern is $a_{i}$. FIGs.~\ref{fig:zeta=01} and \ref{fig:zeta=001} show that there always exist directions of $a_{i} = 0$ if we choose a proper turning time, for both of left and right side of the wormhole.

In order to mathematically check the existence of $a_{i} \rightarrow 0$ direction, we need to look at the equations in detail. The complexified equations are as follows:
\begin{eqnarray}
\ddot{a}_{r} &=& - \frac{8\pi a_{r}}{3} \left( \dot{\phi}_{r}^{2} - \dot{\phi}_{i}^{2} \right) + \frac{16 \pi a_{i}}{3} \dot{\phi}_{r} \dot{\phi}_{i} \mp \frac{8\pi a_{r} V_{0}}{3},\\
\ddot{a}_{i} &=& - \frac{8\pi a_{i}}{3} \left( \dot{\phi}_{r}^{2} - \dot{\phi}_{i}^{2} \right) - \frac{16 \pi a_{r}}{3} \dot{\phi}_{r} \dot{\phi}_{i} \mp \frac{8\pi a_{i} V_{0}}{3},\\
\ddot{\phi}_{r} &=& - \frac{3}{a_{r}^{2} + a_{i}^{2}} \left( \dot{a}_{r} a_{r} \dot{\phi}_{r} + \dot{a}_{r} a_{i} \dot{\phi}_{i} + \dot{a}_{i}a_{i} \dot{\phi}_{r} - \dot{a}_{i} a_{r} \dot{\phi}_{i} \right),\\
\ddot{\phi}_{i} &=& - \frac{3}{a_{r}^{2} + a_{i}^{2}} \left( \dot{a}_{r} a_{r} \dot{\phi}_{i} - \dot{a}_{r} a_{i} \dot{\phi}_{r} + \dot{a}_{i}a_{r} \dot{\phi}_{r} + \dot{a}_{i} a_{i} \dot{\phi}_{i} \right),
\end{eqnarray}
where the upper sign is for the Euclidean signatures and the lower sign is for the Lorentzian signatures. Especially, we can focus on the limit when $|a_{r}| \gg |a_{i}|$ along the Lorentzian time. Then first, the scalar field equations can be simplified as
\begin{eqnarray}
\ddot{\phi}_{r} &\simeq& - \frac{3}{a_{r}} \dot{a}_{r} \dot{\phi}_{r},\\
\ddot{\phi}_{i} &\simeq& - \frac{3}{a_{r}} \dot{a}_{r} \dot{\phi}_{i}.
\end{eqnarray}
Therefore, $\dot{\phi}_{r,i} \simeq C/a_{r}^{3} \simeq C e^{-3 H_{0} t}$, where $H_{0} = \sqrt{8\pi V_{0}/3}$ is the Hubble parameter and $C$ is a constant. Therefore, along the Lorentzian time, the velocities of real and imaginary parts of the scalar field exponentially decrease and the field eventually stops. Then the equations for $a_{r,i}$ are simplified and
\begin{eqnarray}
\ddot{a}_{r} &\simeq& - H_{0}^{2} a_{r},\\
\ddot{a}_{i} &\simeq& - H_{0}^{2} a_{i}.
\end{eqnarray}
Therefore, $a_{r,i} = A e^{-H_{0}t} + B e^{H_{0} t}$, where $A$ and $B$ are constants. For consistency to maintain $|a_{r}| \gg |a_{i}|$, we need to choose initial conditions such that $A = 0$ for $a_{r}$ and $B = 0$ for $a_{i}$. In order to change the initial condition, we can tune the turning time. Finally, we check that the existence of the classicalized direction is self-consistent.
  
In left of FIG.~\ref{fig:zeta=01}, there appears poles or singularities, where $a_{r}$ approaches to zero. Hence, from a pole, there can appear a branch cut. In right of FIG.~\ref{fig:zeta=01}, there is a discontinuous surface of $a_{i} = 0$ (around $\tau=0$) and this is due to poles. One may also interpret these poles as an Ekpyrotic phase \cite{Battarra:2014xoa}.

\begin{figure}
\begin{center}
\includegraphics[scale=0.3]{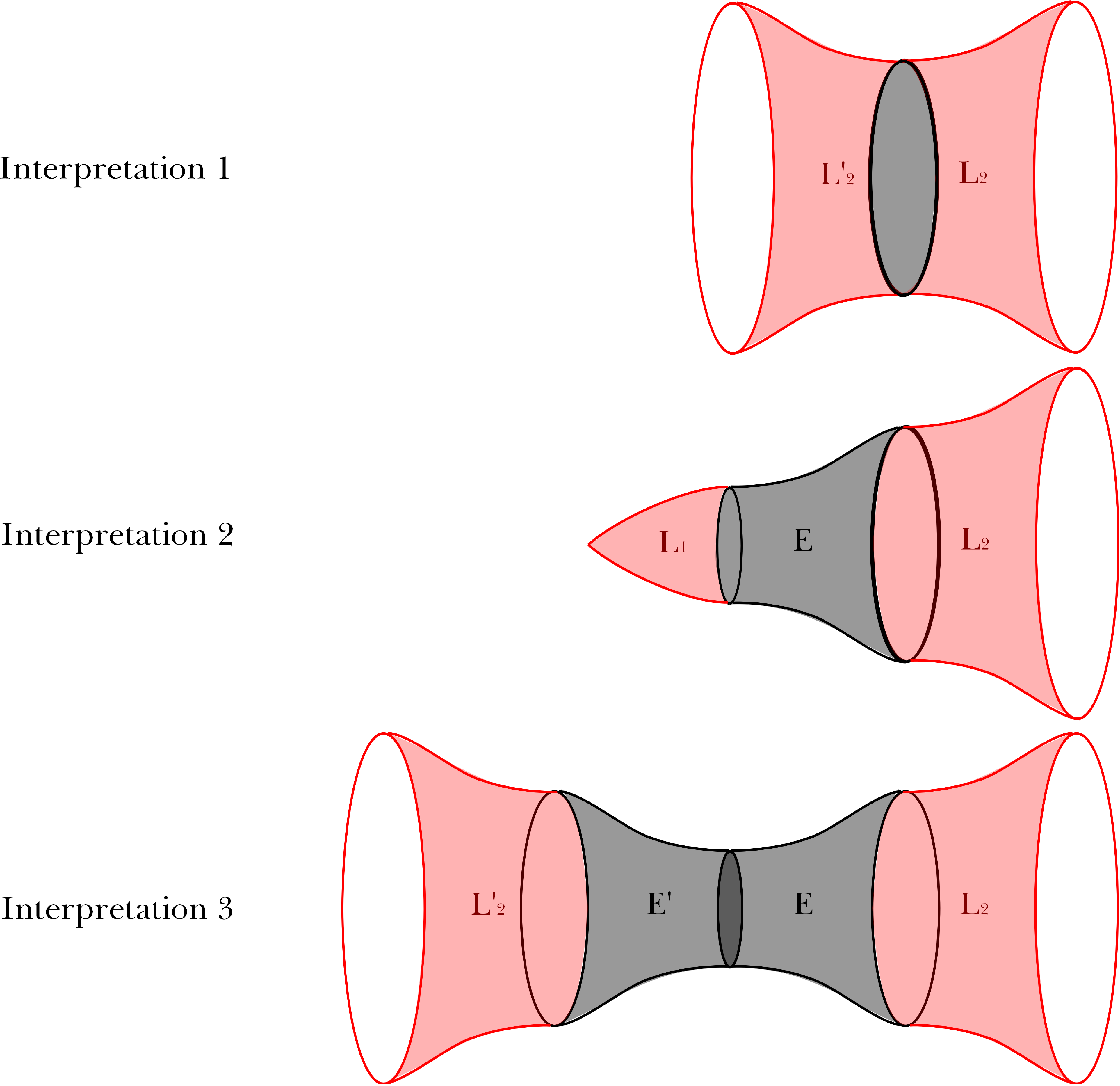}
\caption{\label{fig:bigsol}Upper: A collapsing universe is bounced. In this case, the classical process is dominated, where there can be a quantum contribution (Interpretation 1). Middle: One can interpret that a small universe tunnels to a large universe or one contracting and one expanding universes are created (Interpretation 2). Lower: One can also interpret that two entangled universes are created or a contracting universe is bounced to an expanding universe (Interpretation 3).}
\end{center}
\end{figure}

\begin{figure}
\begin{center}
\includegraphics[scale=0.3]{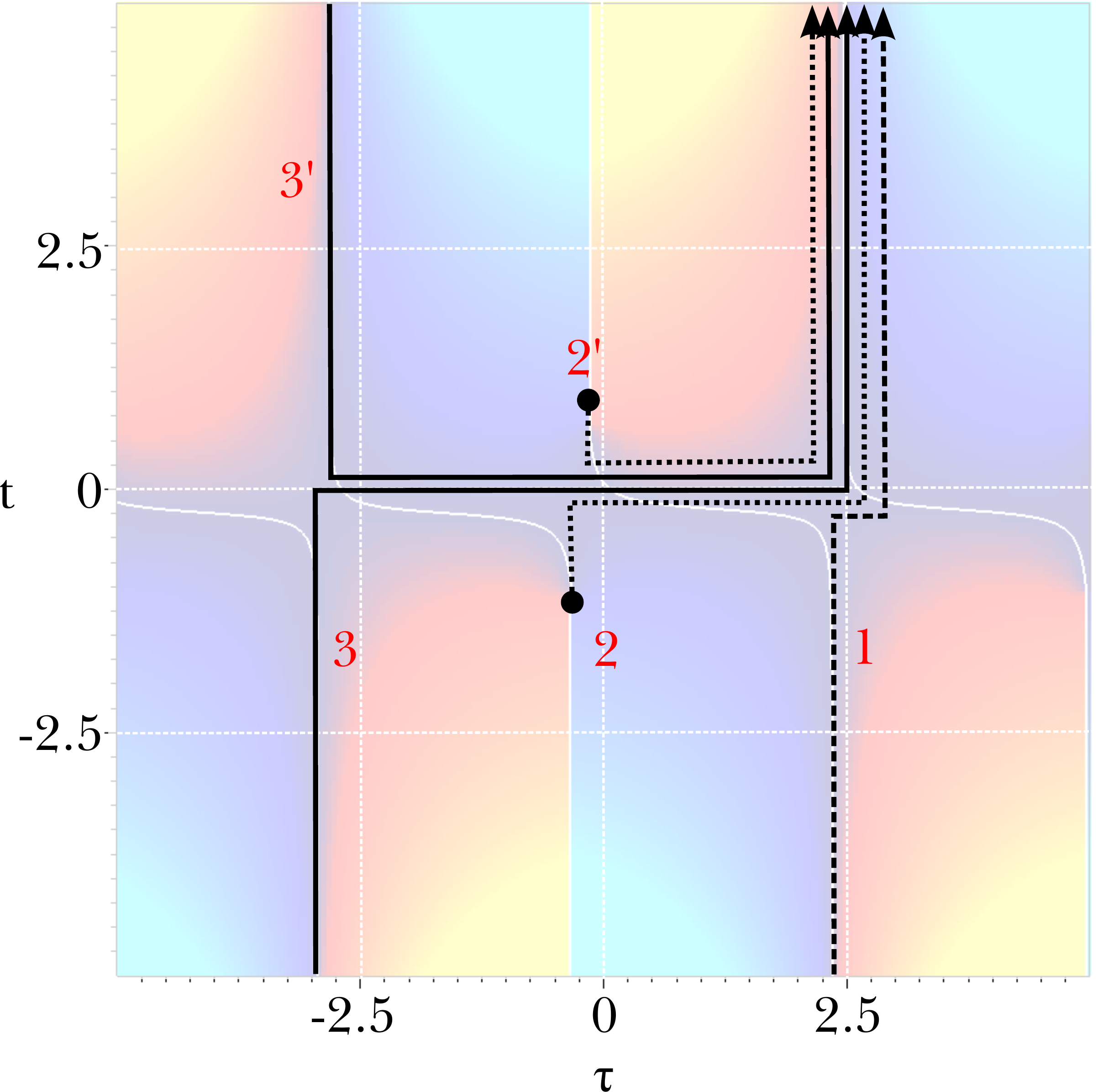}
\caption{\label{fig:contours}Three possible contours and corresponding interpretations (dashed: Interpretation 1, dotted: Interpretation 2, line: Interpretation 3). For Interpretation 2 and 3, the arrows of time can be either one way ($2$, $3$) or two ways ($2'$, $3'$).}
\end{center}
\end{figure}

One interesting point is that there may be no unique classicalized history. Therefore, there can be various interpretations (FIG.~\ref{fig:bigsol}) \cite{Robles-Perez:2013kva}; also, FIG.~\ref{fig:contours} denotes corresponding time contours. There can be basically three interpretations. First, a contracting universe can be classically bounced (Interpretation 1 in FIG.~\ref{fig:bigsol}). This process is classical if there is no scalar field. If there is a contribution of complexified scalar fields, this process may need a small Wick-rotation (see the contour $1$ in FIG.~\ref{fig:contours}). Second, an expanding universe (from a pole, e.g., in left of FIG.~\ref{fig:zeta=01}) can tunnel to an expanding universe (Interpretation $2$ in FIG.~\ref{fig:bigsol}). This means that classically the universe must have been collapsed; however, due to tunneling, it turns to an expanding universe (contour $2$ in FIG.~\ref{fig:contours}). This can be also interpreted that two universes are created from nothing and one is collapsing and the other is expanding (contour $2'$ in FIG.~\ref{fig:contours}). Third, a contracting universe can tunnel to an expanding universe which is intermediated by a Euclidean wormhole (Interpretation 3 in FIG.~\ref{fig:bigsol}). Regarding this, there can be also two interpretations, where one is that a contracting universe tunnels to an expanding universe and hence there is a unique arrow of time (contour $3$ in FIG.~\ref{fig:contours}). However, it is also possible to do an alternative interpretation: two (probably entangled) universes (and two arrows of the time) are created from nothing (contour $3'$ in FIG.~\ref{fig:contours}).

Interpretation 1 and Interpretation 3 can be understood as big bounces, where the former is approximately classical one and the latter is genuinely quantum one. The former is \textit{approximately} (not perfectly) classical, since there is no pure Lorentzian time contour from the collapsing phase to the bouncing phase, as long as there is a contribution of the complexified scalar fields. Therefore, one can interpret that as the universe becomes smaller and smaller, there appears an effective phantom field that helps the big bounce. However, in this case, there is no unique Lorentzian time that connects from the contracting phase to the bouncing phase, and this is one distinct point from usual big bounce scenarios. It is interesting to compare with the big bounce scenario from loop quantum cosmology \cite{Ashtekar:2015dja}: effectively
\begin{eqnarray}
\dot{a}^{2} = \frac{8\pi a^{2} \rho}{3} \left( 1 - \frac{\rho}{\rho_{0}} \right),
\end{eqnarray}
where $\rho$ is the energy density and $\rho_{0}$ is a model dependent constant. Compared to this, our mechanism is not the same as that of loop quantum cosmology.


\section{\label{sec:prob}Probability of wormholes}

The on-shell Euclidean action can be written as follows:
\begin{eqnarray}
S_{\mathrm{E}} = 4 \pi^{2} \int d\tau \left[ a^{3} V - \frac{3}{8\pi} a \right].
\end{eqnarray}
If we integrate over full time period, the probability is $|\Psi|^{2} \propto e^{- \mathrm{Re}\; S_{\mathrm{E}}}$.

After fixing classiclaized contours, we can integrate the Euclidean action and obtain the decay rate. Regarding this, there are two important reference points:
\begin{itemize}
\item[--] \textit{Hawking-Moss instanton} \cite{Hawking:1981fz}: If there is no Euclidean wormhole, then we obtain the Hawking-Moss instanton. The decay rate is
\begin{eqnarray}
S_{\mathrm{E,\;HM}} = - \pi \ell^{2}.
\end{eqnarray}
\item[--] \textit{Approximate solution} (according to Eq.~(\ref{eq:approx})): The on-shell action where the scale factor moves from $a_{\mathrm{max}}$ through $a_{\mathrm{min}}$ to $a_{\mathrm{max}}$ again is
\begin{eqnarray}
S_{\mathrm{E}} \simeq - 3 \pi \int_{a_{\mathrm{min}}}^{a_{\mathrm{max}}} da \frac{a \left(1 - \frac{a^{2}}{\ell^{2}} \right)}{\sqrt{1 - \left( \frac{a_{0}^{4}}{a^{4}} + \frac{a^{2}}{\ell^{2}} \right)}} \label{eq:up}.
\end{eqnarray}
\end{itemize}

\begin{figure}
\begin{center}
\includegraphics[scale=0.75]{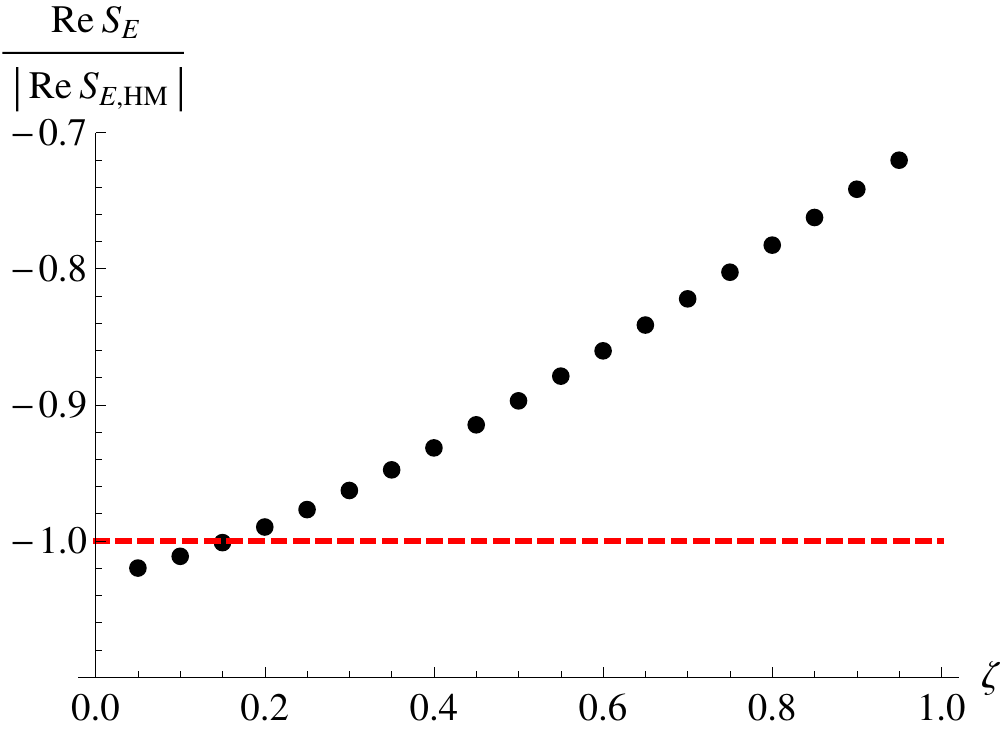}
\includegraphics[scale=0.75]{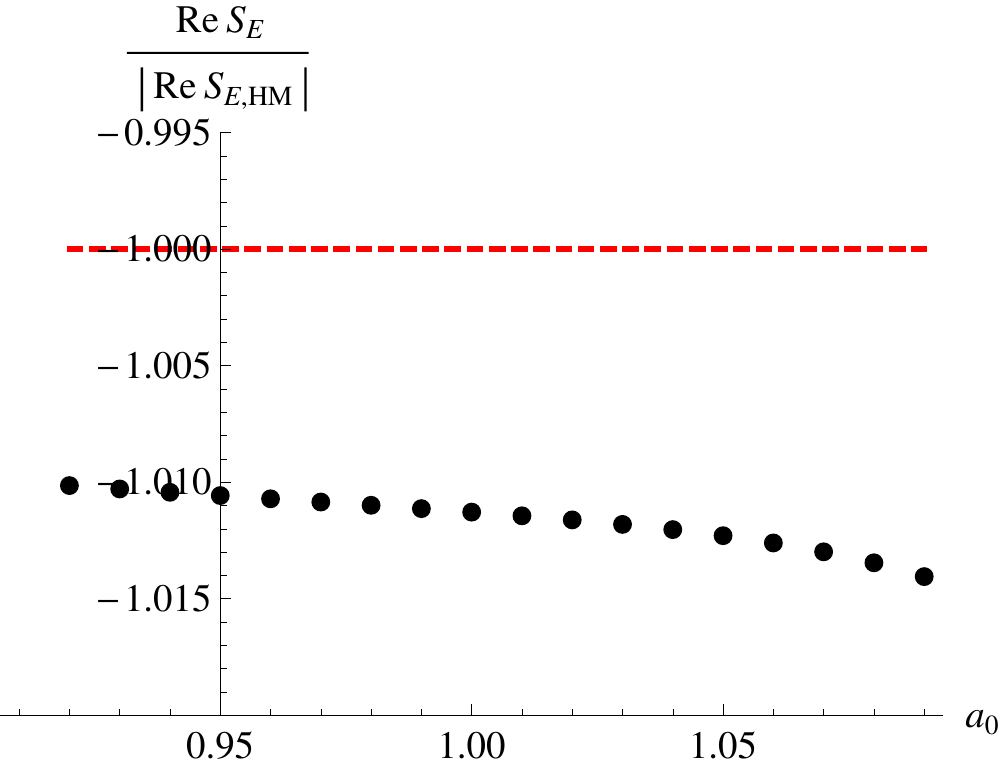}
\caption{\label{fig:action}Left: The Euclidean action by varying $\zeta$ for $a_{0} = 1$ and $\ell = 2$. Right: The Euclidean action by varying $a_{0}$ for $\zeta = 0.1$ and $\ell = 2$. The red dashed line is the Euclidean action for the Hawking-Moss instanton case for $\ell = 2$.}
%\end{center}
%\end{figure}
%\begin{figure}
%\begin{center}
\includegraphics[scale=0.4]{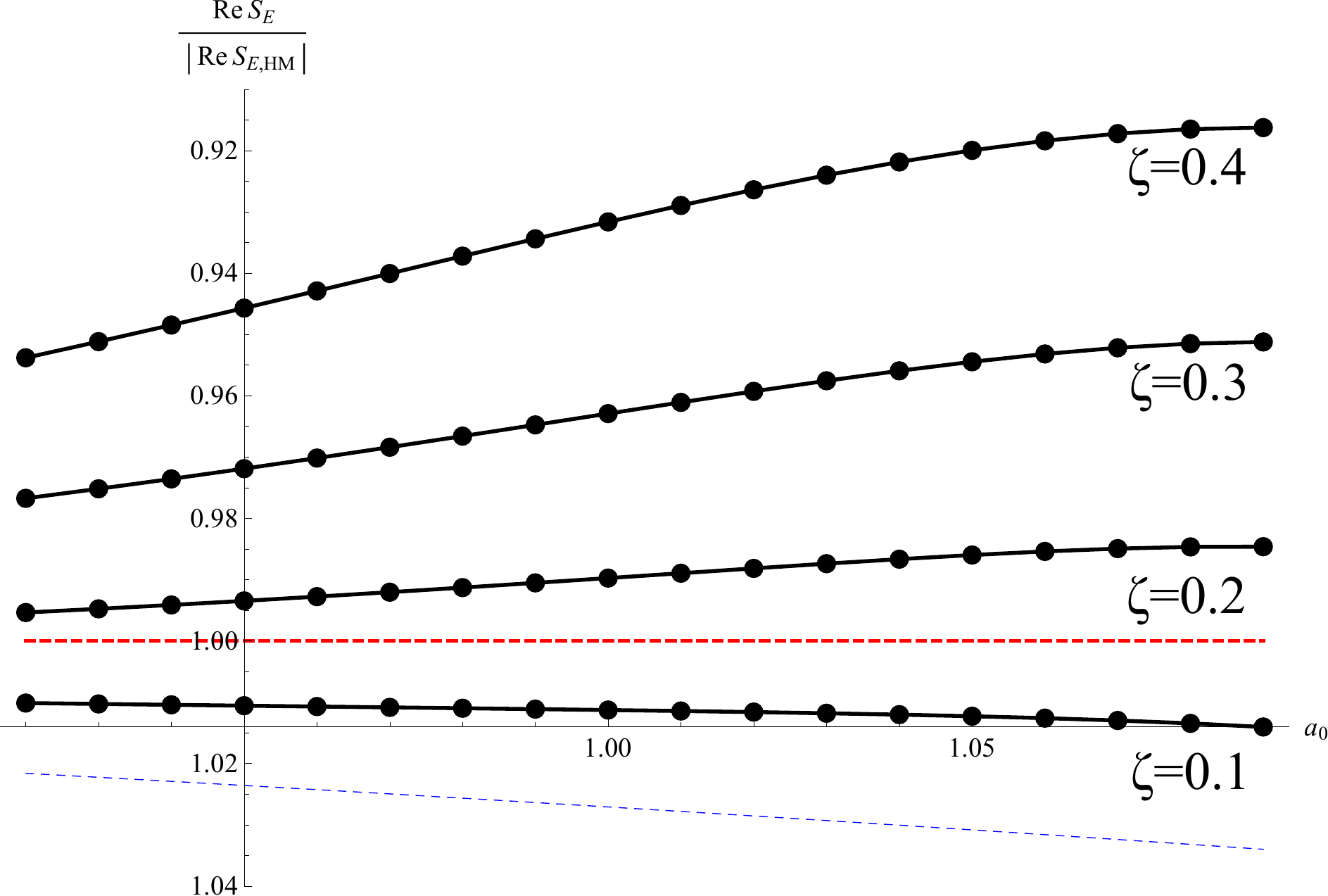}
\includegraphics[scale=0.4]{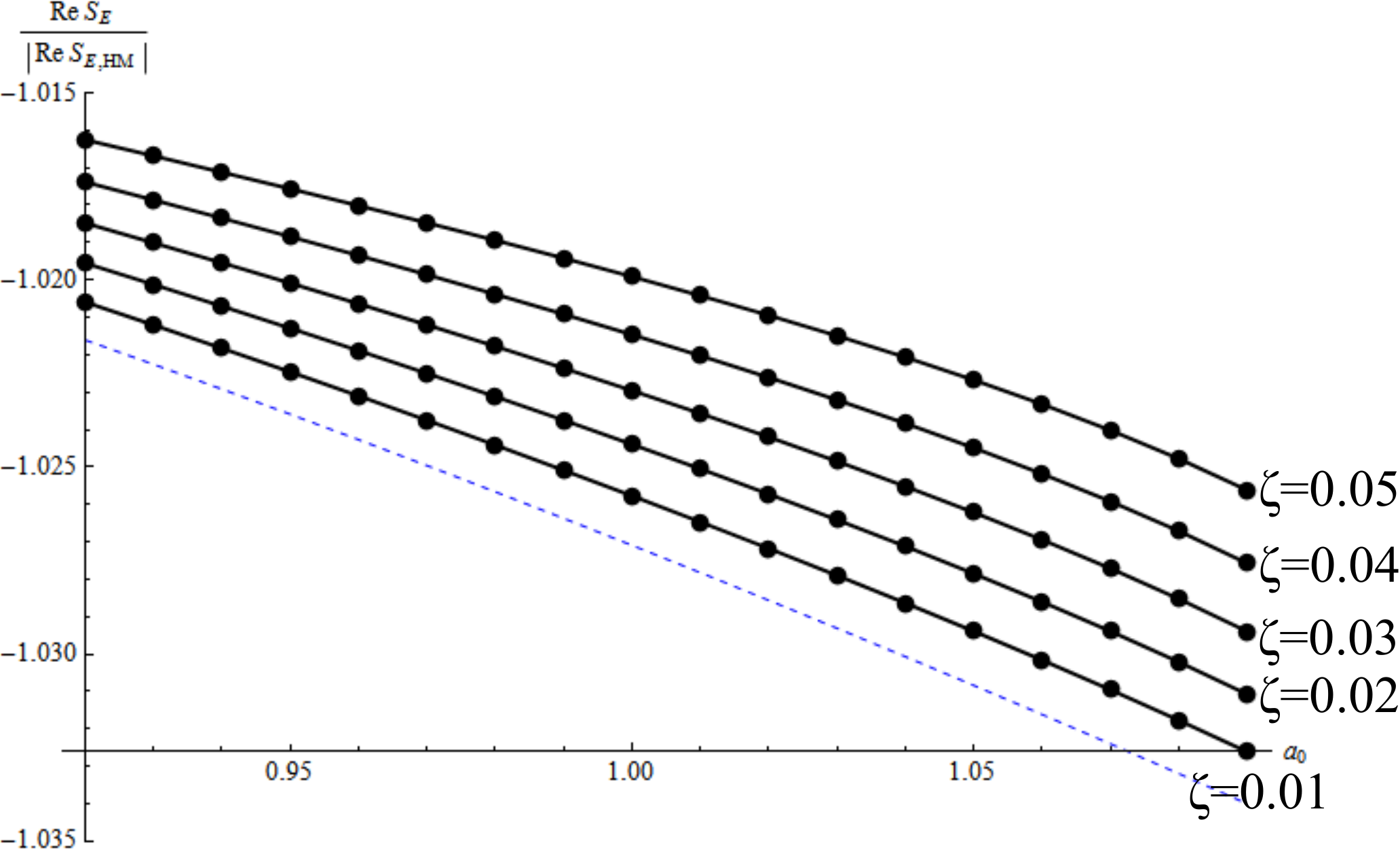}
\caption{\label{fig:probability}The Euclidean action by varying $\zeta$ ($0.1, 0.2, 0.3, 0.4$ (left) and $0.01, 0.02, 0.03, 0.04, 0.05$ (right)) and $a_{0}$ ($\ell = 2$). As $\zeta$ decreases, probabilities approach to the approximate analytic formula Eq.~(\ref{eq:up}) (blue dashed curve). As a reference, the red dashed curve is the case for the Hawking-Moss instantons.}
\end{center}
\end{figure}

Due to the freedom of rescaling, the relative ratio between the action of the solution and that of the Hawking-Moss instanton is physically interesting. As we fix $\ell = 2$ without loss of generality, by varying initial conditions $a_{0}$ and $\zeta$, we can see the probability dependence of various wormholes. FIG.~\ref{fig:action} is the result by varying $\zeta$ (left) and $a_{0}$ (right). We can see a clear dependence that as $\zeta$ decreases, the probability increases and even can be preferred than the Hawking-Moss instantons. FIG.~\ref{fig:probability} shows that as $\zeta$ approaches to zero, the dependence approaches to the blue dashed curve, i.e., the approximate analytic results Eq.~(\ref{eq:up}). Therefore, it is reasonable to conclude that Eq.~(\ref{eq:up}) is the ultimate probability bound of the Euclidean wormholes.

\section{\label{sec:phys}Physical issues}

\subsection{How many arrows of time?: quantum big bounce vs. creation from nothing}

Interpretation 3 can be understood as a quantum bounce. If such a quantum bounce happens using the Hawking-Moss instantons \cite{Hartle:2015bna}, then there is no unique matching between the contracting phase and the bouncing phase; hence, for a given initial contracting phase, the bouncing phase should be a superposition of various universes. On the other hand, if such a quantum bounce happens using the Euclidean wormhole solutions, then there can be a unique connection between the initial phase and the final phase. Note that for very small $\zeta$ cases, the Euclidean wormhole solutions are preferred than the Hawking-Moss instantons. Therefore, it may be possible to conclude that, for the most dominant path, there is a unique continuation of the contracting phase to the bouncing phase, that is consistent with loop quantum cosmology.

Then is there any way to rely on the creation from nothing? In fact, we can interpret Euclidean wormholes by this way, if we choose the time contour as $2'$ or $3'$ of FIG.~\ref{fig:contours}. Then we need to interpret that two universes are created from nothing (and again, this is preferred over Hawking-Moss instantons). %If this is possible, then it challenges us that we can interpret the big bounce scenario in loop quantum cosmology by the same way (creation from nothing and has double time directions).

Can there be any way to distinguish two contours? At the homogeneous level, there may be no good way. However, if we consider perturbations (see also \cite{Ashtekar:2016pqn}), then two interpretations should have different boundary conditions; if there is only a single time direction, then the boundary condition should be given at $t = - \infty$, while if there are double time directions, the boundary condition should be given at $\tau = 0$. In addition to the perturbations, quantum fluctuations as well as their entanglement entropy can give observable differences \cite{Chen:2012bi}. We leave these topics for our future investigations.

In conclusion, we observe that Euclidean wormholes are preferred over Hawking-Moss instantons, and this implies the following issues.
\begin{itemize}
\item[--] If we interpret wormholes based on the assumption of a single arrow of time, then before and after the tunneling, there may be a unique connection between the initial phase and the final phase (rather than superpositions).
\item[--] If we interpret wormholes as based on the assumption of two arrows of time, then this can be considered as the creation of two universes from nothing, and hence universes should be superposed.
\item[--] These two interpretations can be distinguished by observing perturbations or quantum fluctuations, e.g., via cosmic microwave background radiation.
\end{itemize}

\subsection{Interplay between one arrow and two arrows: conspiracy in no-boundary?}

Regarding these wormhole solutions, we have the freedom to interpret them in the following ways. They are either connected uniquely from the contracting phase to the bouncing phase or superposed by different histories with various initial conditions; let us name the former interpretation as \textit{quantum bounce} and the latter as \textit{superposition}.

For Hawking-Moss instantons, traditionally only the superposition interpretation permissible since instantons are compact. On the other hand, for loop quantum cosmology, traditionally only the quantum bounce interpretation is permissible.

At this stage, we may need to expand our thoughts. Perhaps, compact Hawking-Moss instantons may also imply a quantum bounce \cite{Hartle:2015bna}; hence each south pole of Hawking-Moss instantons may have some hidden connections by some unknown reasons. This might not be so strange. As we mentioned in Eq.~(\ref{eq:grndst}), it is possible that the true ground state ($n=0$) of our universe might be a Euclidean wormhole, though this is approximately the same as the Hawking-Moss instanton. If it is the case, then there can be a connection between the contracting phase and the bouncing phase.

In addition, oppositely, it is fair to say that loop quantum cosmology can allow the creation of two universes, where this can be superposed by various initial conditions.

\begin{figure}
\begin{center}
\includegraphics[scale=0.4]{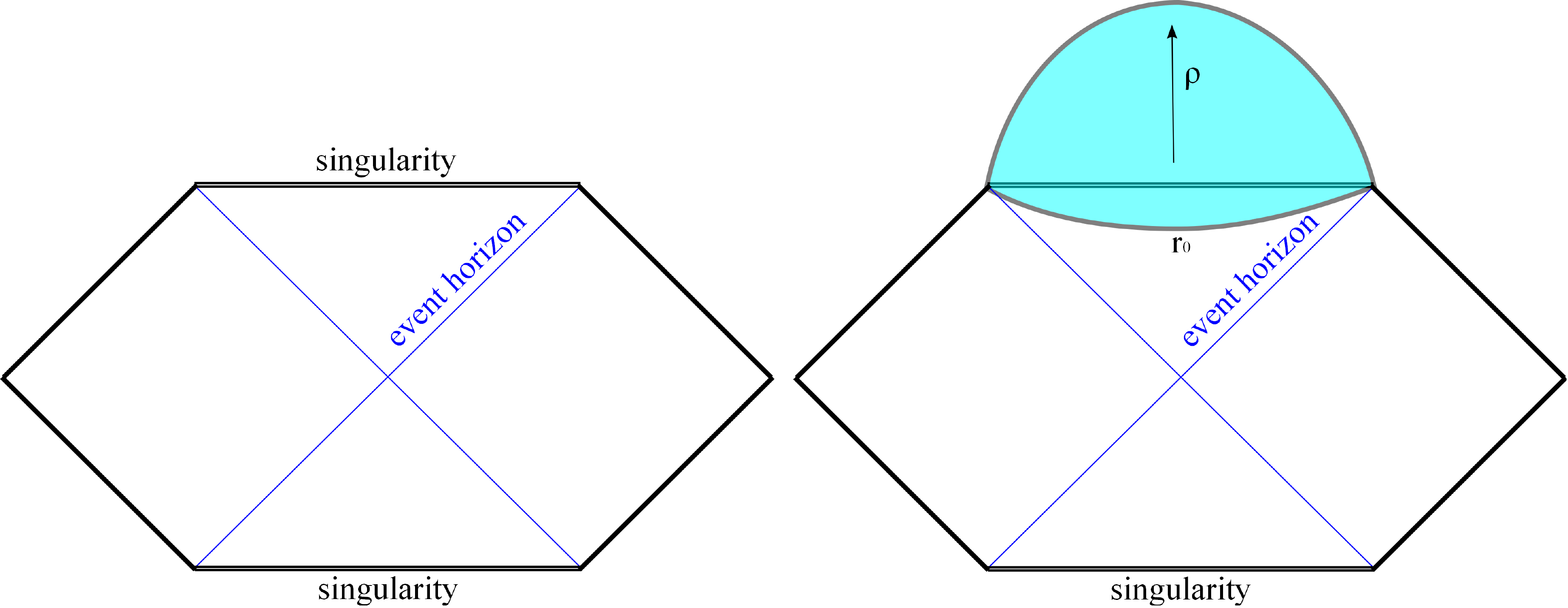}
\caption{\label{fig:wormhole4}Wick-rotation at $r_{0}$ inside the horizon.}
\end{center}
\end{figure}

\begin{figure}
\begin{center}
\includegraphics[scale=0.3]{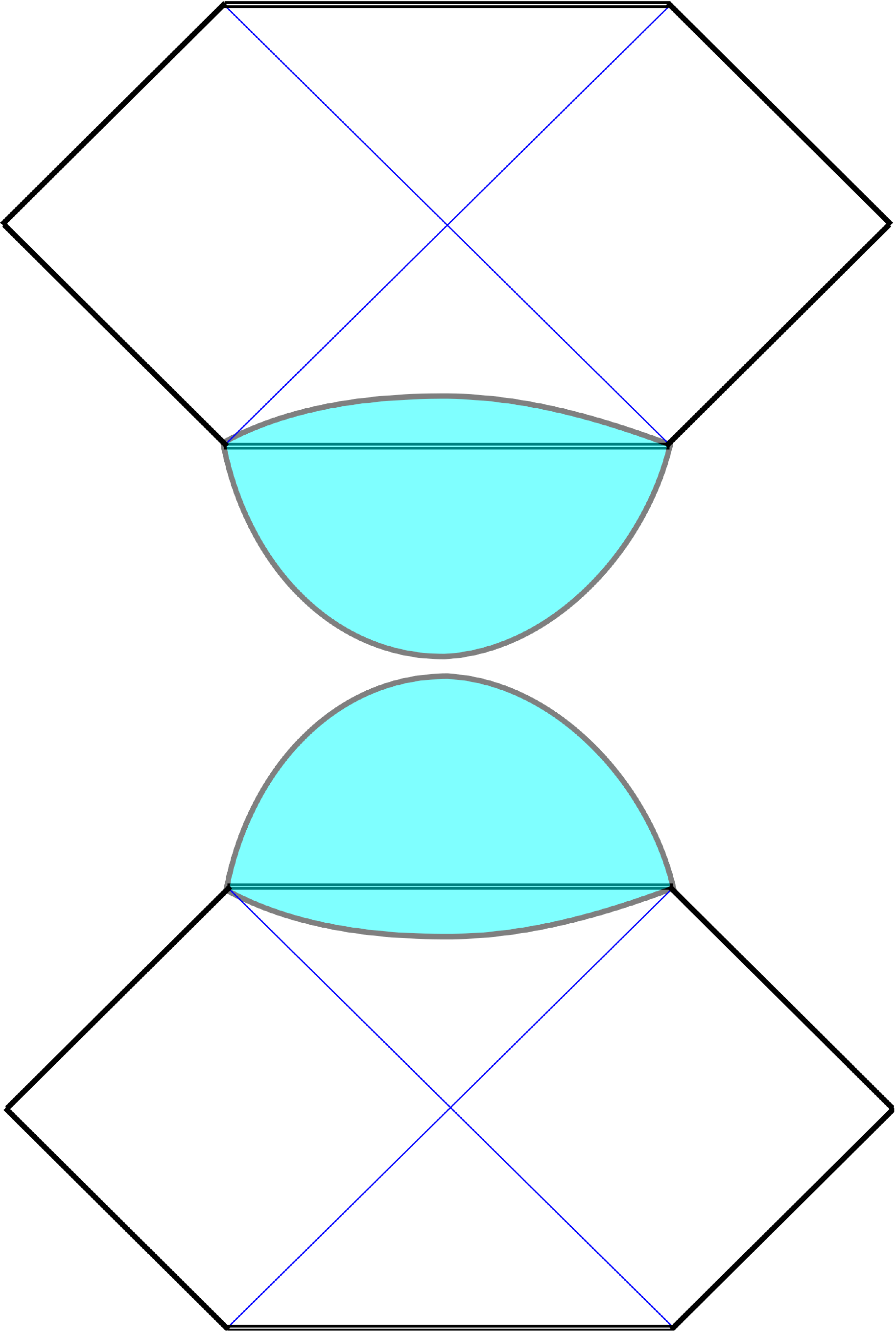}
\caption{\label{fig:wormhole_causalstructure}A possible causal structure for an in-falling observer. There is no principle to connect between the initial manifold and the final manifold. However, if there is a hidden connection between the initial and the final Euclidean manifolds, this can be a reasonable internal structure.}
\end{center}
\end{figure}

One further bold extension of our thought is on black holes. In loop quantum gravity, many people expect a quantum bounce near the black hole singularity. If it really happens, then this can help to understand the causal structure inside the black hole \cite{Haggard:2014rza}. Interestingly, in terms of the Euclidean approach, the similar thing can be explained. For example, the Schwarzschild solution inside the event horizon ($r < 2M$) is
\begin{eqnarray}
ds^{2} = \left( \frac{2M}{r} - 1 \right) dt^{2} - \left( \frac{2M}{r} - 1 \right)^{-1} dr^{2} + r^{2} d\Omega^{2},
\end{eqnarray}
where the time parameter can be Wick-rotated by $r = r_{0} + i\rho$. Then, we obtain
\begin{eqnarray}
ds^{2}_{\mathrm{E}} = \left( \frac{2M}{r_{0} + i\rho} - 1 \right) dt^{2} + \left( \frac{2M}{r_{0} + i\rho} - 1 \right)^{-1} d\rho^{2} + \left(r_{0} + i\rho \right)^{2} d\Omega^{2},
\end{eqnarray}
where in the large $\rho$ limit, the metric is asymptotically classicalized, 
\begin{eqnarray}
ds^{2}_{\mathrm{E}} \rightarrow - \left( dt^{2} + d\rho^{2} + \rho^{2} d\Omega^{2} \right),
\end{eqnarray}
and the geometry approaches Euclidean Minkowski space (FIG.~\ref{fig:wormhole4}).
In this analytic continuation, the Ricci scalar vanishes and hence the Euclidean action may also vanish. If we turn on the cosmological constant (if $r = r_{0} \pm i \rho$), then
\begin{eqnarray}
S_{\mathrm{E}} = \pm \frac{\Lambda}{8\pi}\int \sqrt{+g} dx_{\mathrm{E}}^{4} = \mp \frac{\Lambda}{8\pi} \times \left(\mathrm{Volume}\right).
\end{eqnarray}
Since the volume integration diverges, the probability either vanishes or diverges to infinity. Therefore, unless there is a regularization technique, it is not easy to do meaningful physics using this analytic continuation. However, if there is a conspiracy between two analytic continuations, then there can be a quantum connection between the black hole and the white hole (FIG.~\ref{fig:wormhole_causalstructure}). Of course, this assertion is hypothetical and it requires further investigations.


On the other hand, for the \textit{asymptotic} structure, we may need to interpret it as superpositions and the total causal structure may not be so unique \cite{Sasaki:2014spa,Hartle:2015bna}. These two types of instantons (instantons for inside and outside the horizon) need to be well-combined in order to understand the complete picture of black hole physics. Again, we leave this as a future topic.

\subsection{Loss of unitarity for a semi-classical observer: effective loss of information}

Finally, we briefly comment on the unitarity issue \cite{Hawking:1976ra,Chen:2014jwq}. Would the Euclidean path integral preserve the unitarity of the time evolution? Would the existence of a wormhole spoil this? Previously, people pointed out that the existence of Euclidean wormholes is harmful to unitarity \cite{Hawking:1988ae} or holography \cite{ArkaniHamed:2007js}. However, in almost all of previous works, people only considered wormholes in the Euclidean signatures and this made the problem to be less clear. In contrast, our investigation of wormholes includes Lorentzian analytic continuation, which allows us to define the unitarity issue more explicitly (though limited to de Sitter space and hence difficult to apply holography).

If there is only one arrow of time, then there would be a unique dominant instanton and hence this may not be harmful to the unitarity. However, if there are two arrows of time, then as we can (semi-classical observers) only observe one part of the universe, eventually we will notice that our universe is in a mixed state \cite{Robles-Perez:2013kva}. This does not mean that there is a true loss of information. Instead, a semi-classical observer would detect an apparent loss of information. This is also an example of the effective loss of information \cite{Sasaki:2014spa}. Finally, our conclusion can be summarized as follows: \textit{Euclidean wormholes may be compatible with the unitarity, but can cause an effective loss of information for a semi-classical observer}.

\section{\label{sec:con}Conclusion}

In this paper, we investigated various phenomena of Euclidean wormholes inspired by complexification of fields and metrics. There are some parameters that favors Euclidean wormholes over Hawking-Moss instantons. Moreover, Euclidean wormhole solutions have a family of continuous parameters while the Hawking-Moss instanton corresponds just a single point in the parameter space. Hence, these Euclidean wormholes can be more fundamental than the Hawking-Moss instanton in light of the Euclidean path integral.

Euclidean wormholes can be interpreted in three ways. The first one is the classical bounce, where the process is approximately classical. The second one is quantum tunneling from an expanding universe to a larger universe, where the initial expanding phase can be interpreted as an Ekpyrotic phase; this can be also interpreted as two universes created from nothing where one is expanding and while the other contracting. The third one has two versions; one is a quantum bounce and the other a quantum creation of two universes from nothing.

One interesting observation is that, for a given instanton, there can be two complementary interpretations, where one assumes one arrow of time while the other assumes two arrows of time. If these two interpretations are complementary, then we may imagine a conspiracy or hidden connection between the two (compact or non-compact) instantons; also, we may imagine that a quantum bouncing cosmology can be interpreted as a creation of two universes. This will shed some light on some conceptual tensions between the wave function picture/superposition of histories and quantum big bounce around the singularity, perhaps not only in cosmology but also in black hole physics.

There can be further extensions of our works. We summarize them as follows.
\begin{itemize}
\item[--] In this paper, we just assumed a free scalar field with a constant vacuum energy. However, for realistic applications to inflationary cosmology, we need to introduce more detailed inflaton potential. Therefore, the extension to generic potentials is desirable.
\item[--] Euclidean wormholes can give different quantum effects in terms of perturbations. In particular, Euclidean wormholes can be different from the Bunch-Davies vacuum. Also, the boundary condition of the quantum state can in principle be different, depending on whether there is one arrow of time or two, which can give rise to different results for the quantum state of the universe.
\item[--] We have investigated only de Sitter space, but we may extend the analysis to anti-de Sitter space or asymptotic flat space.
\end{itemize}
In addition to these topics, recently some authors investigated similar objects that connect a contracting phase to an expanding phase \cite{Bramberger:2017cgf}. This paper is closely related to the physics of inflaton as well as other cosmological scenarios, e.g., quantum big bounce. Careful comparison of our interpretations with observable phenomenology is important. We leave these interesting topics for future investigations.

\newpage

\section*{Acknowledgment}

DY is supported by Leung Center for Cosmology and Particle Astrophysics (LeCosPA) of National Taiwan University (103R4000).
YCH would like to thank National Taiwan University and LeCosPA for supporting him as a visitor at Stockholm University and NORDITA where this work was done, and also thank Ingemar Bengtsson and his group for hospitality. PC is supported by Taiwan National Science Council under Project No. NSC 97-2112-M-002-026-MY3 and by Taiwan National Center for Theoretical Sciences (NCTS). PC is in addition supported by US Department of Energy under Contract No. DE-AC03-76SF00515.

\section*{Appendix. Euclidean wormholes in massive gravity}

Starting from the action \cite{Lin:2013aha}
\begin{eqnarray}
S_{\mathrm{E}} = - \int d^{4}x \sqrt{+g} \left(\frac{R}{2} - \Lambda + m_{1}^{2} G^{\mu\nu} f_{\mu\nu} - m_{2}^{2} \left( c_{0} + c_{1} f + c_{2} f^{2} + d_{2} f^{\mu}_{\nu} f^{\nu}_{\mu} + ... \right) \right)
\end{eqnarray}
with the ansatz
\begin{eqnarray}
f_{\mu\nu} = \mathrm{diag} \left[ 0, 1, \sin^{2} \psi, \sin^{2}\psi \sin^{2}\theta \right],
\end{eqnarray}
we can calculate all terms:
\begin{eqnarray}
\sqrt{+g}\frac{R}{2} &=& 2\pi^{2} \times 3a \left(1+\dot{a}^{2}\right),\\
\sqrt{+g}\Lambda &=& 2\pi^{2} \times a^{3} \Lambda,\\
\sqrt{+g}G^{\mu\nu}f_{\mu\nu} &=& 2\pi^{2} \times 3 \frac{- 1 + \dot{a}^{2}}{a},\\
\sqrt{+g}f &=& 2\pi^{2} \times 3a,\\
\sqrt{+g}f^{2} &=& 2\pi^{2} \times \frac{9}{a},\\
\sqrt{+g}f^{\mu}_{\nu} f^{\nu}_{\mu} &=& 2\pi^{2} \times \frac{3}{a}.
\end{eqnarray}

The reduced action is as follows:
\begin{eqnarray}
S_{\mathrm{E}} &=& 2 \pi^{2} \int d\tau \left[-3a\left( 1 + \dot{a}^{2} \right) + a^{3} \Lambda + 3m_{1}^{2} \frac{1-\dot{a}^{2}}{a} + m_{2}^{2} \left( c_{0}a^{3} + 3 c_{1} a + c_{2} \frac{9}{a} + d_{2}\frac{3}{a} \right) \right]\\
&=& 6 \pi^{2} \gamma \int d\tilde{\tau} \left[-a\left( 1 + \dot{a}^{2} \right) + \frac{a^{3}}{3} \Lambda_{\mathrm{eff}} + \frac{\alpha}{a} - m_{1}^{2} \frac{\dot{a}^{2}}{a} \right],
\end{eqnarray}
where (derivations of the second line is for $\tilde{\tau}$)
\begin{eqnarray}
d\tilde{\tau} &=& \gamma d\tau,\\
\alpha &=& \frac{m_{1}^{2} + m_{2}^{2} (3c_{2} + d_{2})}{1-c_{1}m_{2}^{2}},\\
\gamma &=& (1-c_{1}m_{2}^{2})^{1/2},\\
\Lambda_{\mathrm{eff}} &=& \frac{\Lambda+c_{0}m_{2}^{2}}{1-c_{1}m_{2}^{2}}.
\end{eqnarray}
From this action, we can derive the equation of motion:
\begin{eqnarray}
\dot{a}^{2} \left( 1+\frac{m_{1}^{2}}{a^{2}} \right) = 1-\frac{\Lambda_{\mathrm{eff}}}{3} a^{2} - \frac{\alpha}{a^{2}}.
\end{eqnarray}

Finally, the on-shell action is
\begin{eqnarray}
S_{\mathrm{E}} &=& 4 \pi^{2} \gamma \int_{0}^{\tilde{\tau}_{\mathrm{max}}} d\tilde{\tau} \left( a^{3} \Lambda_{\mathrm{eff}} - 3 \left( a + \frac{|\alpha|}{a} \right) \right)\\
&=& -12 \pi^{2} \gamma \int_{0}^{a_{\mathrm{max}}} da \sqrt{\left(1-\frac{\Lambda_{\mathrm{eff}}}{3}a^{2} + \frac{|\alpha|}{a^{2}} \right)\left( a^{2}+m_{1}^{2} \right) }.
\end{eqnarray}
If $a=0$, for regularity, $\dot{a}^{2} = - \alpha/m_{1}^{2}$ is required. Therefore, this requires $\alpha < 0$ for consistency. Then there exists only one bouncing point $a_{\mathrm{max}}$, where
\begin{eqnarray}
a_{\mathrm{max}}^{2} = \frac{3}{\Lambda_{\mathrm{eff}}} \left( \frac{1 + \sqrt{1+4\left|\alpha\right|\Lambda_{\mathrm{eff}}/3}}{2} \right).
\end{eqnarray}
Hence,
\begin{eqnarray}
\tilde{\tau}_{\mathrm{max}} = \int_{0}^{a_{\mathrm{max}}} \sqrt{\frac{a^{2}+m_{1}^{2}}{a^{2}-\Lambda_{\mathrm{eff}}a^{4}/3-\alpha}} da.
\end{eqnarray}
Regarding this, there are two problems: first, $\alpha < 0$ requires that the solution is cosmologically unstable \cite{Lin:2013aha}, and second, the choice of $\dot{a}^{2} = - \alpha/m_{1}^{2}$ makes the action integral singular at $a=0$. Therefore, it is better to say that the only consistent application of this model should include Euclidean wormhole solutions.

\end{document}